\documentclass[10pt,journal,compsoc]{IEEEtran}
\usepackage[T1]{fontenc}
\usepackage{enumitem}
\usepackage{graphicx}
\graphicspath{{fig/}}
\usepackage{subfig}
\usepackage{stfloats}
\usepackage{threeparttable}
\usepackage{tcolorbox}
\usepackage{fancyhdr}
\usepackage{listings}
\usepackage{subfig}
\usepackage{bbding}
\usepackage{pifont}
\usepackage{wasysym}
\usepackage{amssymb}
\usepackage{multirow}
\usepackage[update,prepend]{epstopdf}
\usepackage{tcolorbox}
\usepackage{fancyhdr}
\usepackage{url}
\usepackage{listings}
\usepackage{array}
\usepackage{subfig}
\usepackage{fontawesome}
\usepackage{skak}
\usepackage{letltxmacro}
\usepackage{todonotes}
\usepackage{xspace}
\usepackage{hyperref}
\usepackage{caption}% http://ctan.org/pkg/caption
\usepackage{balance} 
\usepackage{graphicx}
\usepackage{colortbl}
\usepackage{mdwmath}
\usepackage{mdwtab}
\usepackage{pifont}
\usepackage[update,prepend]{epstopdf}
\usepackage{listings}
\usepackage{array}
\usepackage[ruled,vlined,linesnumbered]{algorithm2e}
\SetAlgoSkip{}
\usepackage{times}
\usepackage{enumerate}
\usepackage{booktabs}

\definecolor{mGreen}{rgb}{0,0.6,0}
\definecolor{mGray}{rgb}{0.5,0.5,0.5}
\definecolor{mPurple}{rgb}{0.58,0,0.82}
\definecolor{backgroundColour}{rgb}{0.95,0.95,0.92}
\definecolor{notecolor}{rgb}{0,0.0,0} 

\lstdefinestyle{CStyle}{
	backgroundcolor=\color{backgroundColour},   
	commentstyle=\color{mGreen},
	keywordstyle=\color{magenta},
	numberstyle=\tiny\color{mGray},
	stringstyle=\color{mPurple},
	basicstyle=\footnotesize\ttfamily,
	breakatwhitespace=false, 
	breaklines=true,                 
	captionpos=b,                    
	keepspaces=true,                 
	numbers=left,                    
	numbersep=3pt,                  
	showspaces=false,                
	showstringspaces=false,
	showtabs=false,                  
	tabsize=2,
	aboveskip=2mm,
	belowskip=2mm,
	language=C,
	xleftmargin=0.2in,
	xrightmargin=0.2in
}

\lstdefinestyle{CStyleListing1}{
	backgroundcolor=\color{backgroundColour},   
	commentstyle=\color{mGreen},
	keywordstyle=\color{magenta},
	numberstyle=\tiny\color{mGray},
	stringstyle=\color{mPurple},
	basicstyle=\footnotesize\ttfamily,
	breakatwhitespace=false, 
	breaklines=true,                 
	captionpos=b,                    
	keepspaces=true,                 
	numbers=left,                    
	numbersep=4pt,                  
	showspaces=false,                
	showstringspaces=false,
	showtabs=false,                  
	tabsize=2,
	aboveskip=2mm,
	belowskip=2mm,
	language=C,
	xleftmargin=0.1in,
	xrightmargin=0.05in
}

\newcommand{\EFSA}[1]{{\footnotesize{\emph{e}FSA}}}
\newcommand{\FSA}[1]{{\small\textsc{FSA}}}
\newcommand{\ProjectName}[1]{{\small{\emph{e}FSA}}}

\newcommand{\etal}{\emph{et al. }}
\newcommand{\eg}{\emph{e.g.}}
\newcommand{\ie}{\emph{i.e.}}

\definecolor{clmark}{rgb}{0.0, 0.0, 0.0}

\ifCLASSOPTIONcompsoc
  \usepackage[nocompress]{cite}
\else
  \usepackage{cite}
\fi
\ifCLASSINFOpdf
\else
\fi
\begin{document}
%
% paper title
% Titles are generally capitalized except for words such as a, an, and, as,
% at, but, by, for, in, nor, of, on, or, the, to and up, which are usually
% not capitalized unless they are the first or last word of the title.
% Linebreaks \\ can be used within to get better formatting as desired.
% Do not put math or special symbols in the title.
\title{Checking is Believing: Event-Aware Program Anomaly Detection in Cyber-Physical Systems}
%
%
% author names and IEEE memberships
% note positions of commas and nonbreaking spaces ( ~ ) LaTeX will not break
% a structure at a ~ so this keeps an author's name from being broken across
% two lines.
% use \thanks{} to gain access to the first footnote area
% a separate \thanks must be used for each paragraph as LaTeX2e's \thanks
% was not built to handle multiple paragraphs
%
%
%\IEEEcompsocitemizethanks is a special \thanks that produces the bulleted
% lists the Computer Society journals use for "first footnote" author
% affiliations. Use \IEEEcompsocthanksitem which works much like \item
% for each affiliation group. When not in compsoc mode,
% \IEEEcompsocitemizethanks becomes like \thanks and
% \IEEEcompsocthanksitem becomes a line break with idention. This
% facilitates dual compilation, although admittedly the differences in the
% desired content of \author between the different types of papers makes a
% one-size-fits-all approach a daunting prospect. For instance, compsoc 
% journal papers have the author affiliations above the "Manuscript
% received ..."  text while in non-compsoc journals this is reversed. Sigh.

\author{Long Cheng,~\IEEEmembership{Member,~IEEE,}
		Ke Tian,~%~\IEEEmembership{Fellow,~OSA,}
		Danfeng (Daphne) Yao,~\IEEEmembership{Senior Member,~IEEE,} \\
        Lui Sha,~\IEEEmembership{Fellow,~IEEE,}
        and~Raheem A. Beyah,~\IEEEmembership{Senior Member,~IEEE}% <-this % stops a space
        \iffalse
\IEEEcompsocitemizethanks{
\IEEEcompsocthanksitem Long Cheng is with the School of Computing, Clemson University, USA.\protect
\IEEEcompsocthanksitem Ke Tian is with Microsoft, USA.\protect
\IEEEcompsocthanksitem Danfeng (Daphne) Yao is with the Department of Computer Science, Virginia Tech, VA, USA.\protect
% note need leading \protect in front of \\ to get a newline within \thanks as
% \\ is fragile and will error, could use \hfil\break instead.
%E-mail: see http://www.michaelshell.org/contact.html
\IEEEcompsocthanksitem Lui Sha is with the Department of Computer Science, University of Illinois at Urbana Champaign, Champaign, IL, USA.\protect
\IEEEcompsocthanksitem Raheem A. Beyah is with the School of Electrical and Computer Engineering, Georgia Tech, GA, USA.}
\fi
\thanks{A preliminary version of this work appeared in \cite{Long:2017}.} %and \cite{book:2017:cps}
}

\IEEEtitleabstractindextext{%
\begin{abstract}
Securing cyber-physical systems (CPS) against malicious attacks is of paramount importance because these attacks may cause irreparable damages to physical systems. Recent studies have revealed that control programs running on CPS devices suffer from both control-oriented attacks (\eg, code-injection or code-reuse attacks) and data-oriented attacks (\eg, non-control data attacks). Unfortunately, existing detection mechanisms are insufficient to detect runtime data-oriented exploits, due to the lack of runtime execution semantics checking.
In this work, we propose \emph{Orpheus}, a new security methodology for defending against data-oriented attacks by enf\underline{or}cing cyber-\underline{ph}ysical ex\underline{e}c\underline{u}tion \underline{s}emantics. We first present a general method for reasoning
cyber-physical execution semantics of a control program (\ie, causal dependencies between the physical context/\textcolor{clmark}{event} and program control flows), including the event identification and dependence analysis. As an instantiation of \emph{Orpheus}, we then present a new program behavior model, \ie, the event-aware finite-state automaton (\emph{e}FSA). \emph{e}FSA takes advantage of the event-driven nature of CPS control programs and incorporates event checking in anomaly detection. It detects data-oriented exploits if a specific physical event is missing along with the corresponding event dependent state transition. 
We evaluate our prototype's performance by conducting case studies under data-oriented attacks. Results show that \emph{e}FSA can successfully detect different runtime attacks. Our prototype on Raspberry Pi incurs a low overhead, taking 0.0001s for each state transition integrity checking, and 0.063s$\sim$0.211s for the cyber-physical contextual consistency checking. 
\end{abstract}

% Note that keywords are not normally used for peerreview papers.
\begin{IEEEkeywords}
Cyber-physical systems, Data-oriented attacks, Program anomaly detection, Cyber-physical execution semantics.
\end{IEEEkeywords}

}

% make the title area
\maketitle

% To allow for easy dual compilation without having to reenter the
% abstract/keywords data, the \IEEEtitleabstractindextext text will
% not be used in maketitle, but will appear (i.e., to be "transported")
% here as \IEEEdisplaynontitleabstractindextext when the compsoc 
% or transmag modes are not selected <OR> if conference mode is selected 
% - because all conference papers position the abstract like regular
% papers do.
\IEEEdisplaynontitleabstractindextext
% \IEEEdisplaynontitleabstractindextext has no effect when using
% compsoc or transmag under a non-conference mode.

% For peer review papers, you can put extra information on the cover
% page as needed:
% \ifCLASSOPTIONpeerreview
% \begin{center} \bfseries EDICS Category: 3-BBND \end{center}
% \fi
%
% For peerreview papers, this IEEEtran command inserts a page break and
% creates the second title. It will be ignored for other modes.
\IEEEpeerreviewmaketitle

% Computer Society journal (but not conference!) papers do something unusual
% with the very first section heading (almost always called "Introduction").
% They place it ABOVE the main text! IEEEtran.cls does not automatically do
% this for you, but you can achieve this effect with the provided
% \IEEEraisesectionheading{} command. Note the need to keep any \label that
% is to refer to the section immediately after \section in the above as
% \IEEEraisesectionheading puts \section within a raised box.
% 

\IEEEraisesectionheading{\section{Introduction}\label{sec:intro}}

\IEEEPARstart{C}{yber}-physical systems (CPS) consist of a tightly coupled integration of computational elements and physical components. The computational elements rely on sensors to monitor the physical environment and make control decisions to affect physical processes with feedback loops~\cite{Sharma:2014:MAC}. These systems are widely used to operate critical infrastructure assets, such as electric power grid, oil and natural gas distribution, industry automation, medical devices, automobile systems, and air traffic control~\cite{Sridhar:2012:PIEEE}. In the industrial control domain, CPSs are instantiated as the Industrial Control Systems (ICS), Distributed Control Systems (DCS), or Supervisory Control and Data Acquisition (SCADA) systems~\cite{Cardenas:2011:AAP}. Though CPS and IoT (Internet of Things) are defined with different emphasis and have no standard definitions agreed upon by the research community, they have significant overlaps. In general, CPS emphasizes the tightly coupled integration of computational components and physical world. While IoT has an emphasis on the connection of things with networks. If an IoT system interacts with the physical world via sensors/actuators, we can also classify it as a CPS~\cite{CPSIOT}.

The tight coupling with physical space of CPS brings new security and safety challenges. 
Control programs running on CPS devices monitor physical environments by taking sensory data as input and send control signals that affect physical systems~\cite{Mitchell:2014:SID}. They are critical to the proper operations of CPS, as anomalous program behaviors can have serious consequence, or even cause devastating damages to physical systems~\cite{Abera:CFLAT:2016}. For example, the Stuxnet~\cite{stuxnet:2013} attack allows hackers to compromise the control system of a nuclear power plant and manipulate real-world equipment such as centrifuge rotor speeds, which can be very dangerous. According to ICS-CERT's report~\cite{ICSCERT}, there have been continuously increasing number of cyber attacks targeting critical infrastructure. Therefore, securing CPS against malicious attacks becomes of paramount importance in the prevention of potential damages to physical systems.

Recent studies~\cite{Zimmer:2010:TID, Habibi:2015:ICDCS, Chen:NDSS:2016, Abera:CFLAT:2016, NDSS:2017, DBLP:journals/corr/NymanDZLPAS17} have shown that control programs suffer from a variety of runtime software exploits. These attacks can be broadly classified into two categories: %control-oriented attacks and data-oriented attacks, which are explained below
\begin{itemize}[noitemsep, leftmargin=*]
\item
 {\em Control-oriented attacks} exploit memory corruption vulnerabilities to divert a program's control flows, \eg, malicious code injection~\cite{Francillon:2008:CIA} or code reuse attacks~\cite{Habibi:2015:ICDCS}. Control-oriented attacks in conventional cyber systems (\ie, without cyber-physical interactions) have been well studied~\cite{Shu:2015:raid}. It is possible that existing detection approaches~\cite{Kiriansky:2002:SEV, Abadi:2005:CI, Francillon:2009:DES, Abad:2013:CPSNA, Lu:2015:ACF, Carlini:2015:CBE} are extended to defend against control-oriented attacks in CPS.
\item
{\em Data-oriented attacks} manipulate program's internal data variables without violating its control-flow integrity (CFI), \eg, non-control data attacks~\cite{Chen:2005:NAR}, control-flow bending~\cite{Carlini:2015:CBE}, data-oriented programming~\cite{Hu:2016:SP}. Data-oriented attacks are much more stealthy than attacks against control flows. Because existing CFI-based solutions are rendered defenseless under data-oriented attacks, such threats are particularly alarming. We mainly focus on runtime software exploits, and thus sensor data spoofing attacks~\cite{Wang:2014:ESORICS,Tan:ICCPS:2016} in the physical domain are out of the scope in this work. 
%(\textcolor{red}{such as XXX????})
\end{itemize}
 
Since many control decisions are made based on particular values of data variables in control programs~\cite{Abera:CFLAT:2016}, data-oriented attacks could potentially cause serious harm to physical systems in a stealthy way. We further categorize data-oriented attacks against control programs into two types.
{\em i) \textbf{Attacks on control branch}}, which corrupt critical decision making variables at runtime to execute a \textit{valid-yet-unexpected} control-flow path (\eg, allowing liquid to flow into a tank despite it is full~\cite{Adepu:SwaT:2016} or preventing a blast furnace from being shut down properly as in the recent German steel mill attack~\cite{GermanSteelMill}). 
{\em ii) \textbf{Attacks on control intensity}}, which corrupt sensor data variables to manipulate the amount of control operations, \eg, affecting the number of loop iterations to dispense too 
much drug~\cite{Abera:CFLAT:2016}).

In many instances, CPS can be modeled as \textit{event-driven} control systems~\cite{Derler:2013:CSD,jia2017contexiot}. We refer to events as occurrences of interest that come through the cyber-physical observation process or emitted by other entities (\eg, the remote controller), and trigger the execution of corresponding control actions.
Defending against CPS data-oriented attacks is challenging due to the following reasons. First, data-oriented exploits can achieve attack goals without incurring illegal control flows, thus providing opportunities for attackers to evade all control flow integrity based detections~\cite{Hu:2016:SP}. 
Second, CPS programs normally rely on external sensor events to make control decisions. This physical event-driven nature makes it difficult to predict runtime program behaviors in CPS. Hence, an anomaly detection system needs to check the \textit{runtime integrity} of program behaviors from \textit{both cyber and physical domains}. 
%In addition, the event-driven feature also exposes CPS programs to attack surface in the physical domain, making it vulnerable to false data injection attacks.
%Existing CPS anomaly detection techniques mostly focus on monitoring behaviors of a specific physical system~\cite{Urbina:2016:LIS}, very few studies systematically investigate the problem of data-oriented anomaly detection by monitoring CPS programs' internal behaviors.  
Unfortunately, there exist very few defenses~\cite{CoRR:YoonMCCS15, Abera:CFLAT:2016} and they are ineffective to prevent both attack types due to the lack of runtime execution semantics checking.

\textbf{Goals and Contributions.} 
% The event-driven nature makes it difficult to predict runtime program behaviors through static analysis of the program or model training. 
%Data-oriented attacks in CPS may result in inconsistencies between the physical context and program execution, where executed control-flow paths do not correspond to the observations in the physical environment. 
In this paper, we focus on a new type of runtime attacks that result in inconsistencies between the physical context/event and program execution, where executed control flow paths do not correspond to the observed events. These attacks do not necessarily violate any control flow integrity, so existing techniques based on control flow checking are not effective. We point out the need for an event-aware control-program anomaly detection, which reasons about program behaviors with respect to cyber-physical interactions, \eg, whether or not to open a valve is based on the current ground truth water level of a tank~\cite{Adepu:SwaT:2016}. 
None of existing program anomaly detection solutions~\cite{Shu:2015:raid} has the event-aware detection ability. They cannot detect attacks that cause inconsistencies between program control flow paths and the physical environments.

We address the problem of securing control programs against data-oriented attacks, through enforcing the execution semantics of control programs in the cyber-physical domain. Specifically, our program anomaly detection enforces the consistency among control decisions, values of data variables in control programs, and the physical environments. Our main technical contributions are summarized as follows. %\vspace{-3pt}
\begin{itemize} [noitemsep, leftmargin=*]
	\item  
	We describe a new security methodology, named \emph{Orpheus}, that leverages the event-driven nature in characterizing CPS control program behaviors. We present a general method for reasoning cyber-physical execution semantics of a control program, including the event identification and event dependence analysis\footnote{\textcolor{clmark}{Accompanying materials of this work are available at goo.gl/Wkrdzz}}. 
	\item As an instantiation of \emph{Orpheus}, we present a new event-aware finite-state automaton (\ProjectName{}) model to detect anomalous control program behaviors particularly caused by data-oriented attacks in CPS. By enforcing runtime cyber-physical execution semantics, \ProjectName{} detects subtle data-oriented exploits when physical event are inconsistent with the corresponding event-dependent state transitions. While our exposition of \emph{Orpheus} is on an \FSA{} model at the system call level, the design paradigm of \emph{Orpheus} can be used to augment many existing program behavior models, such as the n-gram model~\cite{Warrender:1999:SP} or HMM model~\cite{Xu:2015:CSF}. 
	%We present a new program behavior model, \ie, \ProjectName{}, for high precision anomaly detection in CPS. \ProjectName{} ensures that the runtime state of a CPS program is coherent with its execution semantics. 
	\item We implement a proof-of-concept prototype on Raspberry Pi platforms, which have emerged as popular devices for building CPS applications~\cite{McLaughlin:2014:NDSS, Abera:CFLAT:2016, Tan:EMSOFT:2016}. Our prototype features: i) A gray-box \FSA{} model that examines the return addresses on the stack when system calls are made, and thus significantly increases the bar for constructing evasive mimicry attacks.
	ii) An LLVM-based event dependence analysis tool to extract event properties from programs and correlate the physical event with runtime program behaviors, which we refer to as {\em cyber-physical execution semantics}. iii) A near-real-time anomaly detector using named pipes, with both local and distributed event verifiers to assess the physical context.
	\item We conduct a thorough evaluation of \ProjectName{}'s performance through real-world CPS case studies. Results show that our approach can successfully detect different runtime data-oriented attacks reproduced in our experiments. Our prototype of the runtime anomaly detector takes $\sim$0.0001s to check each state transition in \ProjectName{} model, $\sim$0.063s for the local event verification, and $\sim$0.211s for the distributed event verification. 
	%\textcolor{clmark}{We also provide a video demo to demonstrate \ProjectName{}'s detection capability\footnote{https://youtu.be/-VEjidSgGIc}}. 
\end{itemize}

%The focus of this paper is on providing new security capabilities by enforcing cyber-physical execution semantics in defending against data-oriented attacks in CPS. 
%Our design is a general approach for event-driven CPS control systems. In Sec.~\ref{sec:limit}, we discuss limitations and in-depth practical deployment issues, including implementation on bare-metal devices and programmable logic controllers (PLCs), possible low overhead tracing with real-time requirements, and program anomaly detection as a service. 

%add new: Attacks on CPS might result in damage to the physical property, as a result of an explosion [16, 56] 

%\textcolor{red}{a running example  [change Fig 4 and 5 to the push-syringe() example] First present eFSA???} 

%\textcolor{red}{release the source code of their system.}

\vspace{-3pt}
\section{Background and Attack Model}\label{sec:threat}

In this section, we introduce the CPS background, and describe the attack model and assumptions of this work.
We use examples to illustrate our new detection capabilities.
%and then present the design overview of \emph{Orpheus} framework.

%and explain the attacks that we will address.
%\textcolor{red}{redraw the figure}

\vspace{-6pt}
\subsection{CPS Background}\label{sec:event}
\begin{figure}[!ht]\vspace{-10pt}
	\centering
	\includegraphics[width=0.88\columnwidth]{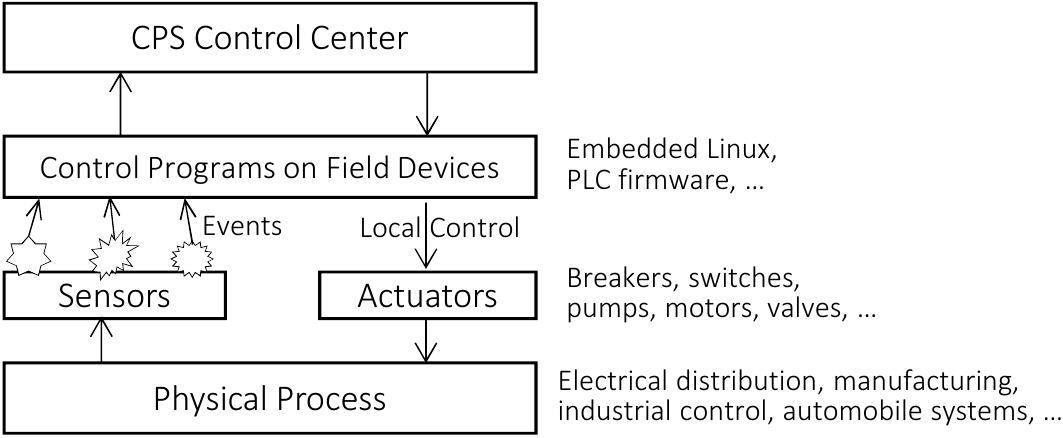}\vspace{-5pt}
	\caption{An abstract view of the event-driven CPS architecture. CPS is exposed with a large attack surface and attacks can be launched across all components in the system.}
	\label{eventmodel}
\end{figure}  \vspace{-3pt}

%Existing CPS anomaly detection approaches mainly monitor behaviors of the physical process. On the contrary, we focus on anomaly detection for CPS programs running on field devices or the central control center in both cyber and physical spaces.

Fig.~\ref{eventmodel} shows an abstract view of the CPS system architecture, which is also in line with the architecture of modern Industrial Control Systems (ICS). In industrial control domain, the control program is often referred to as control logic, and the firmware on PLC (\ie, field device) acts as a kind of operating system~\cite{Physics:NDSS:2017}.
In general, it is composed of the following components: 1) a physical process (\eg, industrial plant or smart home); 2) sensors that measure the physical environment; 3) actuators that trigger physical changes in response to control commands sent by the control program; 4) control programs running on embedded devices that supervise and control physical processes by taking sensory data as input and making local control decisions; 5) a remote control server (which is optional), letting users remotely monitor and control the physical process. CPS communicates with the physical process through sensors and actuators, where physical environments are sensed and events (\eg, coming from the environment or emitted by other entities) are detected, and then actuation tasks are executed through a set of actuators.

Embedded devices (\textit{a.k.a.} field devices) in CPS are situated in the field, where their operating systems are typically embedded Linux/Windows variants~\cite{Schwartz:SP:2014} or PLC firmware~\cite{Physics:NDSS:2017}. Traditionally, embedded control systems were not considered prominent attack targets due to their isolation from potential attack sources. However, the historical isolation has begun to break down as more and more embedded devices are connected to business networks and the Internet in the trend of IoT, making CPS control programs increasingly vulnerable~\cite{Schwartz:SP:2014}.

\subsection{Attack Model and Assumptions}

In this paper, we make the following security assumptions:%\vspace{-3pt}
\begin{itemize}[noitemsep, leftmargin=*]
\item \textit{Capabilities of the adversary.} We assume that the adversary has successfully authenticated CPS field devices (or the control server) under her control to the local network, and is able to launch runtime software exploits which may be unknown or known but unpatched at the time of intrusion. We are not concerned how attackers gained entry into the devices and launch different attacks, but focus on uncovering abnormal program execution behaviors after that~\cite{Lu:2015:ACF}. This is a typical assumption in existing anomaly detection works. 
%We mainly focus on runtime software exploits, and thus sensor data spoofing attacks in the physical domain~\cite{Tan:ICCPS:2016} are out of the scope of this work. 

\item \textit{CPS platform.} We assume the initial state (\ie, the training stage) of the application is trustworthy, which is a general requirement of most behavior-based intrusion detection systems~\cite{CoRR:YoonMCCS15}. We also assume the runtime monitoring module (\textcolor{clmark}{running on the host}) is trusted and cannot be disabled or modified. This assumption is reasonable because it can be achieved by isolating the anomaly detector (monitoring module) from the untrusted target program with hardware security support such as Intel's TrustLite or ARM's TrustZone~\cite{Abera:CFLAT:2016}. At the time of detection, the user space is partially or fully compromised, but the operating system space has not been fully penetrated yet, and thus it is still trusted~\cite{Zimmer:2010:TID}. 
\item \textit{Our focus.} We focus our investigation on runtime software exploits, and thus sensor data spoofing attacks in the physical domain~\cite{Tan:ICCPS:2016} are out of the scope. We assume sensor measurements are trustable. We limit our attention to data-oriented attacks that involve \textit{changes of system call usage}. Other data-related attacks that do not impact observable program behavior patterns (\eg, modification of non-decision making variables) are beyond the scope of this work. System call can be used as an ideal signal for detecting potential intrusions, since a compromised program can generally cause damage to the victim system only by exploiting system calls~\cite{Mutz:2006:ASC}.
Despite system call based monitoring is widely used for detecting compromised programs, we aim at developing a CPS-specific anomaly detection system by augmenting an existing program behavior model with physical context awareness. 
\end{itemize}

\subsection{New Detection Capabilities}

Our new detection capability is detecting data-oriented attacks in CPS control programs, including hijacked for/while-loops or conditional branches. These stealthy attacks alter the underlying control program's behaviors without tampering control-flow graphs (CFGs).
\textcolor{clmark}{We illustrate our new detection capabilities using a smart syringe pump as an example~\footnote{\scriptsize https://hackaday.io/project/1838-open-syringe-pump}. 
The control program reads humidity sensor values as well as takes remote user commands, and translates the input values/commands into control signals to its actuator. Partial code is shown in Listing~\ref{lst:syringe}. Suppose a stack buffer overflow vulnerability exists in the \texttt{recvRemoteCommand()} function (line 3). When the vulnerable function is invoked, an attacker
is able to corrupt the sensor variables (\eg, \texttt{pressure}, \texttt{temperature}, and \texttt{humidity}) in the program.} Our approach reasons about control programs' behaviors w.r.t physical environments, and is able to detect the following attacks: 
\begin{itemize}[noitemsep, leftmargin=*]
	\item
	{\em Attacking control branch.} An attack affecting the code in lines 5 and 7 of Listing~\ref{lst:syringe} may trigger \texttt{push-syringe} or \texttt{pull-syringe} regardless of physical events or remote requests. It corrupts control variables that result in event function \verb1push_event1 or \verb1pull_event1 returning \verb1True1. Such an attack leads to unintended but valid control flows.
	\item
	{\em Attacking control intensity.} An attack may directly or indirectly corrupt a local state variable (\eg, \texttt{steps} in line 18 of Listing~\ref{lst:syringe}) that controls the amount of liquid to dispense by the pump. Such an attack may cause the syringe to overpump than what is necessary for the physical system. Range-based anomaly detection would not work, as the overwritten variable may still be within the permitted range but incompatible with the current physical context. Such an attack (\ie, manipulating the control loop iterations) does not violate the program's CFG either. 
\end{itemize}

%basicstyle=\scriptsize

\begin{lstlisting}[caption={\textcolor{clmark}{Examples of data-oriented attacks in a simplified smart syringe pump application. An attacker could purposely (a) trigger control actions by manipulating the return value of \texttt{push\_event} or \texttt{pull\_event}, and (b) manipulate the number of loop iterations in \texttt{push-syringe} without violating the control program's CFG.}},label={lst:syringe}, style=CStyleListing1, escapeinside={(*}{*)}]
void loop(...) {
(*~\enskip\enskip*)readSensors(&pressure,&temperature,&humidity);
(*\faBug*) recvRemoteCommand();/*buffer overflow vulnerability*/
  ...
(*\withattack*) if(push_event()==True)/*Attack control branch*/
			push_syringe();
(*\withattack*) else if (pull_event()==True) 
			pull_syringe();
}
bool push_event() {
	//decide whether push_event is triggered
	if(humidity>HUMIDITY_THRESHOLD) 
		return True;
	return False;
}
void push_syringe() {
	//calculate the steps value
(*\withattack*)(*\enskip*)steps=humidity-HUMIDITY_THRESHOLD; 
	for(int i=0; i<steps; i++){/*Attack control intensity*/
		digitalWrite(motorStepPin,HIGH);
		delayMicroseconds(usDelay);
		digitalWrite(motorStepPin,LOW);
	}
}

\end{lstlisting}

\iffalse
\begin{figure}[th]
	\centering %\vspace{-6pt}
	\includegraphics[width=0.7\columnwidth]{ProblemExample.pdf}\vspace{-3pt}
	\caption{Two examples of data-oriented software exploits in a real-world CPS application. An attacker could purposely (a) trigger control actions by manipulating the return value of \texttt{Push\_Event} or \texttt{Pull\_Event}, and (b) manipulate the number of loop iterations in \texttt{push-syringe} without violating the control program's CFG.} %\vspace{-5pt}
	\label{Example}
\end{figure} %
\fi 

Existing solutions cannot detect these attacks, as the detection does not incorporate events and cannot reason about program behaviors w.r.t. physical environments. C-FLAT~\cite{Abera:CFLAT:2016}, which is based on the attestation of control flows and a finite number of permitted execution patterns, cannot fully detect these attacks. 
Similarly, recent frequency- and co-occurrence-based anomaly detection approaches (\eg, global trace analysis~\cite{Shu:2015:USP} and system call frequency distribution (SCFD)~\cite{CoRR:YoonMCCS15}) cannot detect such either type of attacks, as their analyses do not model runtime cyber-physical context dependencies.

\subsection{Definition of Events}
Without loss of generality, we define two types of events in control programs: \emph{binary events} and \emph{control-intensity events}. \textcolor{clmark}{In this work, the physical context refers to these physical events that trigger a particular execution path in a CPS program.} %\vspace{-3pt}
\begin{itemize}[noitemsep, leftmargin=*]
	\item Binary events return either \verb1True1 or \verb1False1, which are defined in terms of pre-specified status changes of physical environments and provide notifications to the control program (\eg, \verb1push_event1 or \verb1pull_event1 in Listing~\ref{lst:syringe}). 
	Note that though sensor values such as temperature or humidity have continuous attributes that would lead to a large input space, binary events have a binary outcome which indicates a pre-specified status change is triggered. Such events are commonly pre-defined and used in CPS/IoT's trigger-action programming ("if, then") model~\cite{Ur:2014:PTP, jia2017contexiot}. 
	\item \textcolor{clmark}{Control-intensity events correspond to the sensor-driven control actions within a for/while loop, \eg, sensor values affect the amount of control operations of \texttt{push-syringe} in Listing~\ref{lst:syringe}. We consider each loop iteration as a discrete event.
		It is challenging to identify control-intensity events since they are not explicitly declared in control programs. We present a general event identification method in Sec.~\ref{sec:design:epfsa:event}}.
\end{itemize}

\begin{figure*}[!ht]% \vspace{-9pt}
	\centering
	\includegraphics[width=0.65\textwidth]{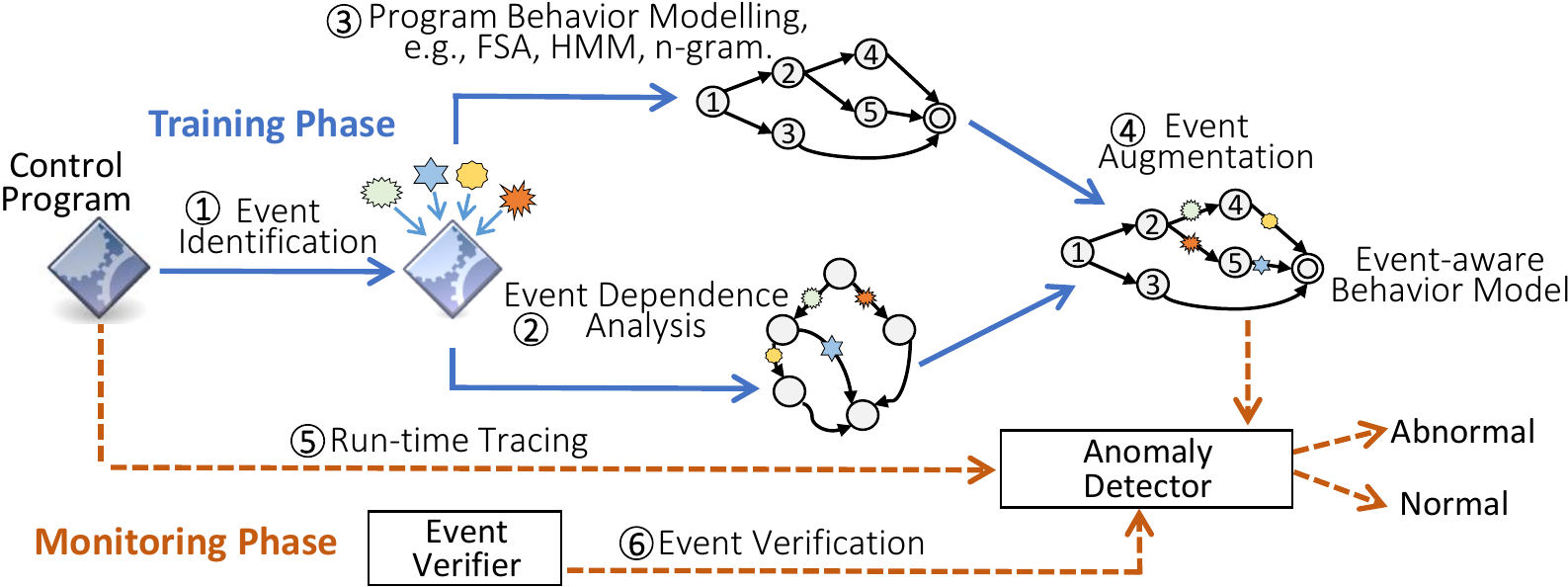}%\vspace{-5pt}
	\caption{Workflow of \emph{Orpheus} event-aware anomaly detection framework, which augments an existing program behavior model with  cyber-physical contextual integrity.}\vspace{-12pt}
	\label{workflow}
\end{figure*}%

\section{\emph{Orpheus} Anomaly Detection Framework}\label{sec:overview}

\subsection{Motivation}\label{sec:designoverview:motivation}

% C memory corruption still there,  --> When attacks happen, cause inconsistency between program runtime behavior and the program execution semantics,

\textcolor{clmark}{Runtime software attacks by exploiting memory corruption vulnerabilities constitute a major attack vector against CPS~\cite{2007:SP:Clements}\cite{Getahun:2018}. This is because, low-level memory-unsafe languages (\eg, C/C++) are widely used in embedded systems for speed performance purposes. As launching control-oriented attacks become increasingly difficult due to many deployed defenses against control-flow hijacking, data-oriented attacks are considered an appealing attack technique for system compromise.}

\textcolor{clmark}{
Data-oriented attacks can purposely change the underlying CPS program behaviors and drive the system to unexpected states in a stealthy way, and thus posing a serious security threat to CPS. From the example in Listing~\ref{lst:syringe}, we observe that runtime control flows of CPS program are dependent on the external physical context. A data-oriented attack could lead to an inconsistency between the physical context and program control flow.
This motivates us to leverage the \emph{intrinsic physical context dependency} in CPS control programs as a channel to detect anomalous program behavior in CPS. Our key idea is to enforce physical context constraints over existing program behavior models, and check the consistency between runtime program behavior and external execution semantics.}

%Such vulnerabilities could lead to serious security problems in CPS and would remain an unsolved problem for a long time. 

%whether a specific physical event associated with this event-dependent state transition is observed in the physical domain.

%physical event-driven nature makes it difficult to predict runtime program behaviors in CPS.

%Motivation. 

\subsection{Design Overview}\label{sec:designoverview}

Fig.~\ref{workflow} shows the workflow of \emph{Orpheus} event-aware anomaly detection framework, which is a learning-based program anomaly detection and composed of two stages: training/learning (where program behavior models are built based on normal program traces) and monitoring/testing (where a new trace is compared against the model built in the training phase). 
%for constructing the \EFSA{} program behavior model in our design. 
In particular, to capture the cyber-physical context dependency of control programs, the training stage in \emph{Orpheus} encompasses both static program analysis and dynamic profiling.
% where the goal of static analysis is to extract the event dependent program slices. 

There are four main steps in the training phase. In step~\ding{172} (Sec.~\ref{sec:design:epfsa:event}), \emph{Orpheus} identifies both binary events and control-intensity events involved in the control program.
In step~\ding{173} (Sec.~\ref{sec:analysis}), it performs the program dependency analysis to generate event-annotated CFG, which identifies the instructions/statements associated with binary events, and control-intensity loops associated with control-intensity events. In step~\ding{174} (Sec.~\ref{sec:design:efsa}), \emph{Orpheus} constructs the normal program behavior model either based on static analysis or dynamic profiling, which we refer to as a basic program behavior model in \emph{Orpheus}. The next step~\ding{175} (Sec.~\ref{sec:design:efsa2}) is important. It augments the basic model with event constraints and obtains the event-aware program behavior model. 

%By augmenting the event-driven information over the basic \FSA{}, we generate our event-aware \FSA{} (\ie, \EFSA{}) for CPS program behavior modeling (\ding{175}).
%Given an event-annotated CFG, we are able to identify the event-driven system call sequences. 

%We first identify both binary events and non-binary events involved in the control program (\ding{182}). After that, we perform the event dependence analysis to generate an event-annotated CFG (\ding{183}), which identifies event triggered instructions/statements of the program (corresponding to binary events), and control intensity loops (corresponding to non-binary events). Then, we construct the basic finite-state automaton (\FSA{}) model based on dynamic profiling (\ding{184}). Given an event-annotated CFG, we are able to identify the event-driven system call sequences. To detect control intensity anomalies, we conduct a control intensity analysis and associate the results with corresponding non-binary events. By augmenting the event-driven information over the basic \FSA{}, we generate our event-aware \FSA{} (\ie, \EFSA{}) for control program behavior modeling (\ding{185}). he runtime monitoring module is trusted and cannot be disabled or modified

Steps~\ding{176} and \ding{177} are the monitoring phase (Sec.~\ref{sec:design:eventverify}). In step~\ding{176}, the anomaly detector (which can be located in the secure world in ARM TrustZone to provide a trusted execution environment for trace collection~\cite{Abera:CFLAT:2016}\cite{OEI:Sun:2018}) monitors the program's execution and collects runtime traces. The basic program behavior model normally aims at detecting control-oriented attacks. Our main contribution lies in the event awareness enhancement on top of the basic model. In the monitoring phase, whenever an event-dependent control-flow path is encountered in step~\ding{177}, the event verifier checks the consistency between runtime behavior and program execution semantics, \eg, whether a specific physical event associated with an event-dependent control-flow path is observed in the physical domain. An anomaly is marked if there exists any deviation from the normal behavior model, or a mismatch between the physical context and program control-flow path. 

%\textcolor{red}{the authors should implicitly mention where the runtime monitoring module is installed. Now it is not clear why the existing trusted computing techniques can be really applied.} 

%\textcolor{red}{In Section 2.5, the authors should provide the intuition behind their design before describing the steps. It seems some descriptions are not consistent. For example, what are the relations among training, monitoring, and monitoring phases? In Section 2.5, I am not sure if it is necessary that the authors mention different program behavior analysis techniques. Section 4 presents the design of the eFSA technique. Can you directly say that Orpheus uses eFSA? I understand that the authors want to claim that their design is generic. However, it may not be good that we ask users to choose the technique.}

\subsection{Program Behavior Model Choices}\label{sec:designoverview:choices}

\textcolor{clmark}{Program behavior modeling has been an active research topic over the past decade and various models have been proposed for legacy applications~\cite{Shu:2015:raid}. Existing models can be classified into two categories: i) local model (\eg, n-gram model~\cite{Warrender:1999:SP}, hidden markov model (HMM) based approach~\cite{Xu:2015:CSF}, finite-state automaton (\FSA{}) model~\cite{Sekar:2001:FAM}); and ii) long-range model 
(\eg, frequency distribution based models~\cite{Shu:2015:USP, Shu:2017:LPB, CoRR:YoonMCCS15}). 
Local anomaly detection inspects short-range segments of program execution traces to detect anomalies such as control-flow violations. Long-range anomaly detection examines longer system behaviors (\eg, a complete program behavior instance) than the local anomaly detection, which can detect frequency anomalies. 
Among these models, system-call based monitoring is widely used for detecting compromised programs, in comparison to library/function-calls.}
 
\textcolor{clmark}{N-gram based model defines the normal program behavior for a process by using short sequences of system-calls. In the training phase, it builds a n-gram database by sliding a window of length $n$ over the system-call traces of normal program executions. An anomaly is detected if a new n-gram is observed in the testing sequence (\ie, test for membership in the database). Although short-range ordering of system-calls have a high probability of being perturbed when abnormal activities occur, it is vulnerable to mimicry attacks~\cite{Wagner:2002:MAH}.
An attacker may insert a malicious code, issuing system-calls accepted by a normal behavior model yet still carries out the same malicious action. 
Instead of using short sequences and being limited by length, state-based models use finite state machine (FSM) to express possible sequences, where the PC information (\ie, program counters which are the return addresses of system-calls) are often used in these models.} 

% can be applied to most of the aforementioned program behavior models
\textcolor{clmark}{The design paradigm of \emph{Orpheus} is to augment physical event constraints on top of an existing program behavior model. For example, automaton/state-based models can be enhanced with event checking on event-dependent state transitions. For the n-gram model~\cite{Warrender:1999:SP}, we identify event-dependent n-grams in the training phase and apply the event checking when observing any event-dependent n-gram in the monitoring phase. In addition, control-flow integrity~\cite{Abadi:2005:CI, Tan:EMSOFT:2016} can also be augmented with event checking before executing control flow transfers. We leave the option of choosing the underlying basic program behavior model open to system developers, which may depend on the specific resource constraints on CPS platforms. For example, compared with the n-gram model, tracing PC information in FSA/HMM during program execution incurs an extra runtime overhead, which we will demonstrate in Sec.~\ref{sec:eva}. We instantiate the \emph{Orpheus} framework using the \FSA{} model~\cite{Sekar:2001:FAM} in Sec.~\ref{sec:design}.}

%Warrender~\etal\cite{Warrender:1999:SP} presented the comparison of four different program behavior models, including simple enumeration of sequences, sequence frequency-based (\ie, n-gram), rule induction-based data mining approach, and Hidden Markov Model (HMM). Sekar~\etal\cite{Sekar:2001:FAM} proposed to construct an \FSA{} via dynamic learning from past traces.
%Recently, Xu~\etal\cite{Xu:2015:CSF} proposed a probabilistic HMM-based control flow model representing the expected call sequences of the program for anomaly detection. Shu~\etal\cite{Shu:2015:USP, Shu:2017:LPB} proposed an anomaly detection approach with two-stage machine learning algorithms for large-scale program behavioral modeling.

%Different from these program behavior models for legacy applications, in this paper, we propose a customized \EFSA{} model for detecting anomalies in CPS. Existing program anomaly detection models mainly focus on control flow integrity checking, and thus can not detect runtime data-oriented attacks. \EFSA{} focuses on detecting data-oriented exploits, and the capability for detecting control-oriented exploits inherits from the underlying \FSA{}. 

%\subsection{Security Policies}\label{sec:designoverview:policy}

\section{Reasoning About Cyber-Physical Execution Semantics}\label{sec:reason}

In this section, we present a general method for reasoning about cyber-physical execution semantics of a control program through static analysis,
including the event identification and dependence analysis. 

%These two models complement each other to detect a broad class of attacks that penetrate and modify the execution behaviors of CPS programs.
%\vspace{-8pt}

%In what follows, we first 
%Then, we describe details about how to build the \EFSA{} model.

\subsection{Event Identification}\label{sec:design:epfsa:event}
%why need event identification?
In order to discover the triggering relationship between external events and internal program control flows, we first identify what events are involved in a control program. For pre-defined binary events, it is not difficult to identify these events (\eg, given event functions declared in an event library or header file, we scan the source code or executable binary). The main challenge is to identify: i) control intensity events/loops, and ii) non-predefined binary events. Our LLVM-based~\cite{LLVM} event identification algorithm can automatically extract these events and only requires knowledge of sensor-reading APIs and actuation APIs on the embedded system. They are pre-specified sources and sinks\footnote{\textcolor{clmark}{Source and sink are terms in a dataflow/taint analysis. The source is where data comes from, and the sink is where it ends in a program~\cite{Schwartz:2010:YEW}.}} in our static analysis.

\begin{algorithm}[!h]
	\vspace{8pt}
	\SetKwFunction{Event}{Event}
	\SetKwFunction{ConstructPDG}{ConstructPDG}
	\SetKwFunction{ConstructDDGraph}{ConstructDDGraph}
	\SetKwFunction{ConstructCDGraph}{ConstructCDGraph}
	\SetKwFunction{getLoopBrSet}{getLoopBranchSet}
	\SetKwFunction{getNextInst}{getNextInst}
	\SetKwFunction{DataDependenceTree}{DataDependenceTree}
	\SetKwFunction{DataControlDependenceTree}{DataControlDependenceTree}
	\SetKwFunction{ForwardCDSet}{ForwardControlDependence}
	\SetKwFunction{BackwardDDSet}{BackwardDataDependence}
	\SetKwFunction{FALSE}{FALSE}
	\SetKwFunction{TRUE}{TRUE}
	\small
	\textbf{Input:} Program $P$; Sensor-reading API set $API_{sens}$; Actuation API set $API_{actu}$\\
	\textbf{Output:} Control-intensity event/loop set $E_{ci}$\
	\SetAlgoLined
	\BlankLine
	\footnotesize
	$E_{ci}\gets\emptyset$;
		
	$G_{pdg}$ = \ConstructPDG{$P$}  \emph{/*construct the program dependence graph*/\;}
	
	%cdGraph = \ConstructCDGraph{$P$}  \emph{/*construct the control dependence graph at the basic block level*/\;}
	%including both data and control dependencies. 
	
	LoopBrSet = \getLoopBrSet{$P$} \emph{/*get all the conditional branch instructions with loops*/\;}
	
	\For{BranchInst=\getNextInst{LoopBranchSet}} {
		$S_{bdd}$ = \BackwardDDSet{$G_{pdg}$, BranchInst}\; \emph{/*Backward data dependent statements on $BranchInst$*/\;}	
		$S_{fcd}$ = \ForwardCDSet{$G_{pdg}$, BranchInst}\; \emph{/*Forward control dependent statements on $BranchInst$*/\;}
		\uIf{ ($S_{bdd}\cap API_{sens}\neq\emptyset$ \& $S_{fcd}\cap API_{actu}\neq\emptyset$) } 
		{$E_{ci}$= $E_{ci}\cup$ \Event{BranchInst,$S_{bdd}$,$S_{fcd}$}\;}
		
	}
	\caption{Identifying control-intensity events}
	%\addtocontents{loa}{\vspace{-2.75\baselineskip}} 
	\label{designn:alg1}    
	%\addtocontents{loa}{\vskip 9pt}
\end{algorithm}

According to the definition of a control-intensity event in Sec.~\ref{sec:event}, it contains a loop statement (\eg, for/while loop) in which sensor values affect the amount of control operations. Our key idea is to search for a loop statement that is data-dependent on any sensor-reading API, and at least an actuation API is control-dependent on this loop statement. The search is performed through backward data dependence analysis and forward control dependence analysis. Algorithm~\ref{designn:alg1} describes our static analysis for identifying control-intensity events. We first obtain the LLVM Intermediate Representation (IR) of a control program $P$ using the \texttt{Clang} compiler~\cite{LLVM}, and construct the program dependence graph (PDG), including both data and control dependencies (Line 4). The control dependence graph is at the basic block level\footnote{In program analysis, a basic block is a linear sequence of instructions containing no branches except at the very end.}, while the data dependency graph is at the granularity of instructions. Then, we obtain all conditional branch instructions with loops, by searching the conditional "br" instruction, which takes a single "i1" value and two "label" values in LLVM IR (Line 5). For each conditional branch with a loop, we conduct the backward inter-procedural dataflow analysis to find any prior data dependence on sensor-reading APIs (Line 7). Then, we conduct the forward inter-procedural control-dependence analysis on the true branch of the conditional instruction to find actuation APIs, \eg, APIs in WiringPi library or functions writing GPIO pins~\cite{WiringPi} (Line 9). If a loop statement is data-dependent on external sensor data, and triggers a certain control action, we identify a control-intensity event/loop (Line 11). In each iteration, we record the identified control-intensity event and control intensity loop (Line 12), which is the output of the event identification process.

\begin{figure}[!h] %\vspace{-10pt}
	\centering
	\includegraphics[width=0.95\columnwidth]{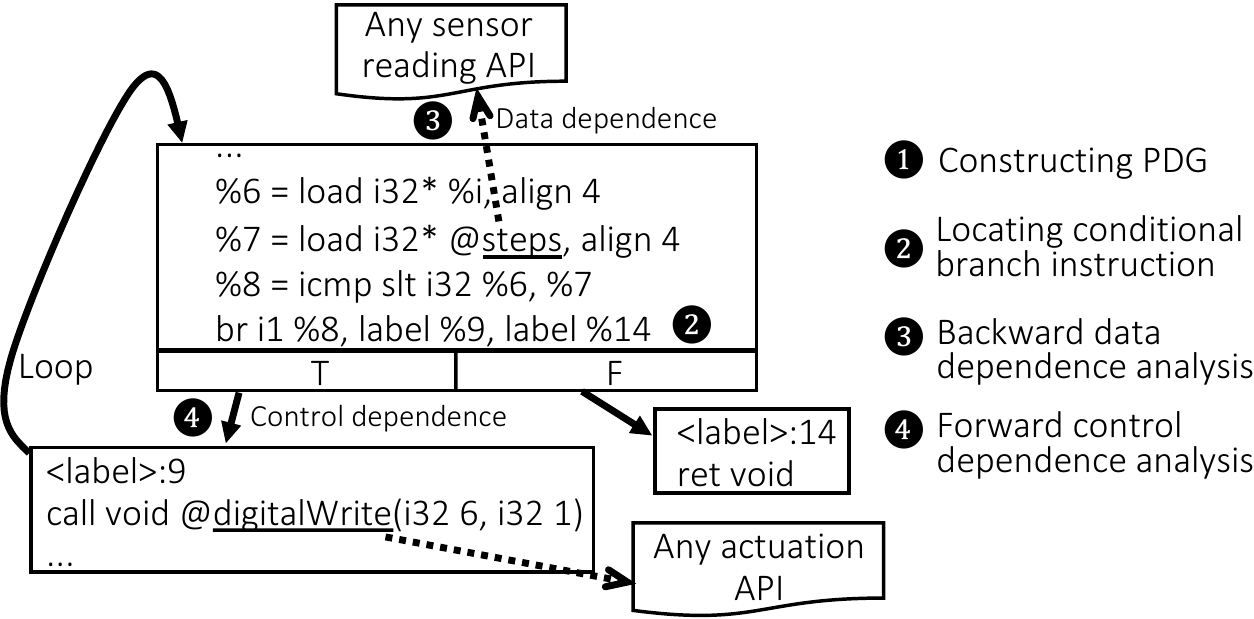} %\vspace{-10pt}
	\caption{\textcolor{clmark}{An example of identifying control-intensity events}}  
	\label{EventIdentification} %\vspace{-10pt}
\end{figure}

\textcolor{clmark}{
A more specific example of our event identification is illustrated in Fig.~\ref{EventIdentification} corresponding to the C-based control program in Listing~\ref{lst:syringe}. 
The figure shows a control-intensity event/loop represented by LLVM IR after the data dependence and control dependence analysis (\ding{182}). We then locate a conditional branch instruction with a loop (\ding{183}). This conditional branch uses the variable \texttt{steps}, which is data dependent on a sensor-reading API (\ding{184}). On its true branch, we find an actuation API \texttt{digitalWrite} and thus we identify the loop as a control-intensity event (\ding{185}).}

%Finally, we record the search results for the next event dependence analysis (\ding{186}). 

We also design a similar procedure for identifying non-pre-defined binary events. An example of such event is when the humidity exceeds a user-designated value, an event predicate returns \verb1True1. In this procedure, we search for the conditional branch either "br" or "switch" instruction without a loop, and then perform the same data/control dependence analysis. In particular, we need to analyze both true and false branches of a "br" instruction, because both branches may contain control actions and we also consider the not-happening case (\ie, the branch without triggering any control action) as an implicit event. 

%*** [Long, add pseudocode ***, [*** pseudocode code for detection ***]]

\subsection{Event Dependence Analysis}\label{sec:analysis} 
Our event dependence analysis generates an event-annotated CFG, \ie, approximating the set of statements/instructions that connect events and their triggered actions. 
During the event identification, we identify individual events that are involved in a control program. 
For the control-intensity event/loop, we directly associate it with the whole loop that contains the sensor-driven control action. A challenge arises when dealing with nested binary events. 
We address the nested events challenge using a bottom-up approach for recursive searching for event dependencies. 

Algorithm~\ref{designn:alg2} describes our event dependence analysis for nested binary events. Given a binary-event triggered basic block $BB_{eta}$, we backward traverse all its control dependent blocks until reaching the root in a recursive manner, and extract corresponding branch labels (\ie, \verb1True1 or \verb1False1). In the recursive function \texttt{FindEventDependence} (Line 5), once we find a basic block on which $BB_{cur}$ is control dependent (Line 7), we check whether it contains any external event (Line 9). If yes, we add this event together with its branch label to $E_{b}$ (Line 10). The condition $E_{b}\cap E_{tmp}=\emptyset$ avoids potential loops when including new events into $E_{b}$. Then, we recursively search any upstream event that $BB_{cur}$ depends on (Line 12).

\begin{algorithm}
	\vspace{8pt}
	\SetKwFunction{FindEveDependence}{FindEventDependence}  
	\SetKwFunction{getNextBB}{getNextBB}
	\SetKwFunction{GetEvent}{extractEvent}
	\SetKwFunction{FALSE}{FALSE}
	\SetKwFunction{TRUE}{TRUE}
	\SetKwProg{Fn}{Function}{}{}
	\small
	\textbf{Input:} Event-triggered basic block $BB_{eta}$; Control flow graph $G_{cfg}$ of program $P$; \\
	\textbf{Output:} $E_{b}$: events that trigger the execution of $BB_{eta}$\
	\SetAlgoLined
	\BlankLine
		\footnotesize
	$E_{b}\gets\emptyset$\;
    $BB_{cur}$ = $BB_{eta}$\;
	\Fn{\FindEveDependence($BB_{cur}$, $G_{cfg}$, $E_{b}$)}{
		\For{$BB_{tmp}$= \getNextBB{$G_{cfg}$}} {
			
			\uIf{ ($BB_{tmp}.toid$ == $BB_{cur}$) } 
			{
				$E_{tmp}$=\GetEvent{$BB_{tmp}$} \;
				
				\uIf{ $E_{tmp}\neq\emptyset$ \& $E_{b}\cap E_{tmp}=\emptyset$ } 
				{
					$E_{b}$= $E_{b}\cup E_{tmp}$\;			
					$BB_{cur}= BB_{tmp}$\;
					\FindEveDependence($BB_{cur}$, $G_{cfg}$, $E_{b}$)\;
				}
			}	
			
		}
		return\;
			
	}	
	\caption{Event dependence analysis for binary events}
	%\addtocontents{loa}{\vspace{-2.75\baselineskip}} 
	\label{designn:alg2}    
	%\addtocontents{loa}{\vskip 9pt}

\end{algorithm}	

\begin{figure}[h] %\vspace{-8pt}
	\centering
	\includegraphics[width=0.99\columnwidth]{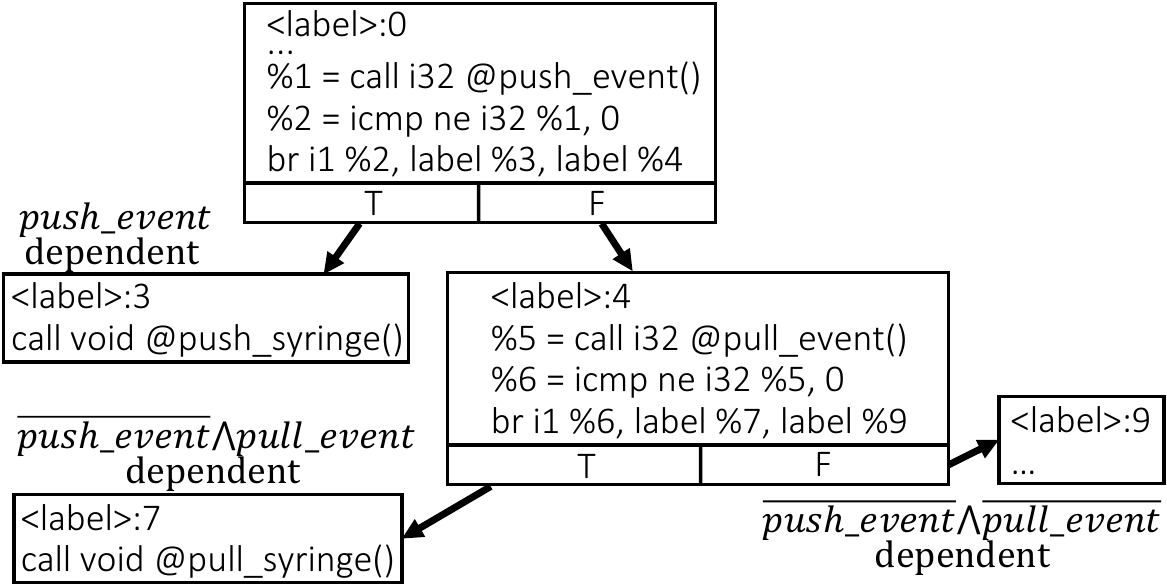} %\vspace{-8pt}
	\caption{\textcolor{clmark}{Event dependence analysis for nested events}}  
	\label{NestedEvent} %\vspace{-9pt}
\end{figure}

\textcolor{clmark}{
Fig.~\ref{NestedEvent} illustrates an example of our event dependence analysis corresponding to Listing~\ref{lst:syringe}. Block 7 (\ie, the basic block with label 7) is control dependent on Block 4 in the \texttt{True} branch of \texttt{pull\_event} (called true-control-dependent). By backward traversing the control dependence graph, we find Block 4 is further false-control-dependent on \texttt{push\_event} in Block 0. Then, we know Block 7 is control dependent on a composite event [\texttt{$\overline{push\_event}\wedge pull\_event$}]. In this example, we also find Blocks 3 control dependent on \texttt{push\_event}, and
Block 9 is control dependent on [\texttt{$\overline{push\_event}\wedge \overline{pull\_event}$}]. We finally identify three event-dependent basic blocks, and obtain the corresponding event-annotated CFG.}
%which are the outputs of the event dependence analysis. 

In addition to the static analysis approach, an alternative for event dependence analysis is using dynamic slicing~\cite{Zhang:2003:PDS}, which identifies statements triggered by a particular event during multiple rounds of program executions. 
It is worth mentioning that our event identification and dependence analysis is a general approach for reasoning cyber-physical execution semantics, independent of specific program anomaly detection models.

\section{\textit{e}FSA: an Instantiation of \emph{Orpheus}}\label{sec:design}

In this section, we describe details about how to build the event-aware finite-state automaton (\ie, \EFSA{}) model, a system call level \FSA{}-based instantiation of the \emph{Orpheus} framework. \EFSA{} captures the event-driven feature of CPS programs to detect evasive attacks. 

%In this and the next few sections, we describe a specific \FSA{}-based anomaly detection model, 

\subsection{Formal Description of \textit{e}FSA}\label{sec:design:efsa}

We construct the finite-state automaton (\FSA{})~\cite{Sekar:2001:FAM} model, which is based on tracing the system calls and program counters (PC) made by a control program under normal execution. Each distinct PC (\ie, the return address of a system call) value indicates a different state of the \FSA{}, so that invocation of same system calls from different places can be differentiated. Each system call corresponds to a state transition. Since the constructed \FSA{} uses memory address information (\ie, PC values) in modeling program behaviors (called the gray-box model), it is more resistant to mimicry attacks than other program models~\cite{Gao:2004:GPT, Shu:2015:raid}. 

\begin{figure}[h]  %\vspace{-6pt}
	\centering
	\includegraphics[width=0.99\columnwidth]{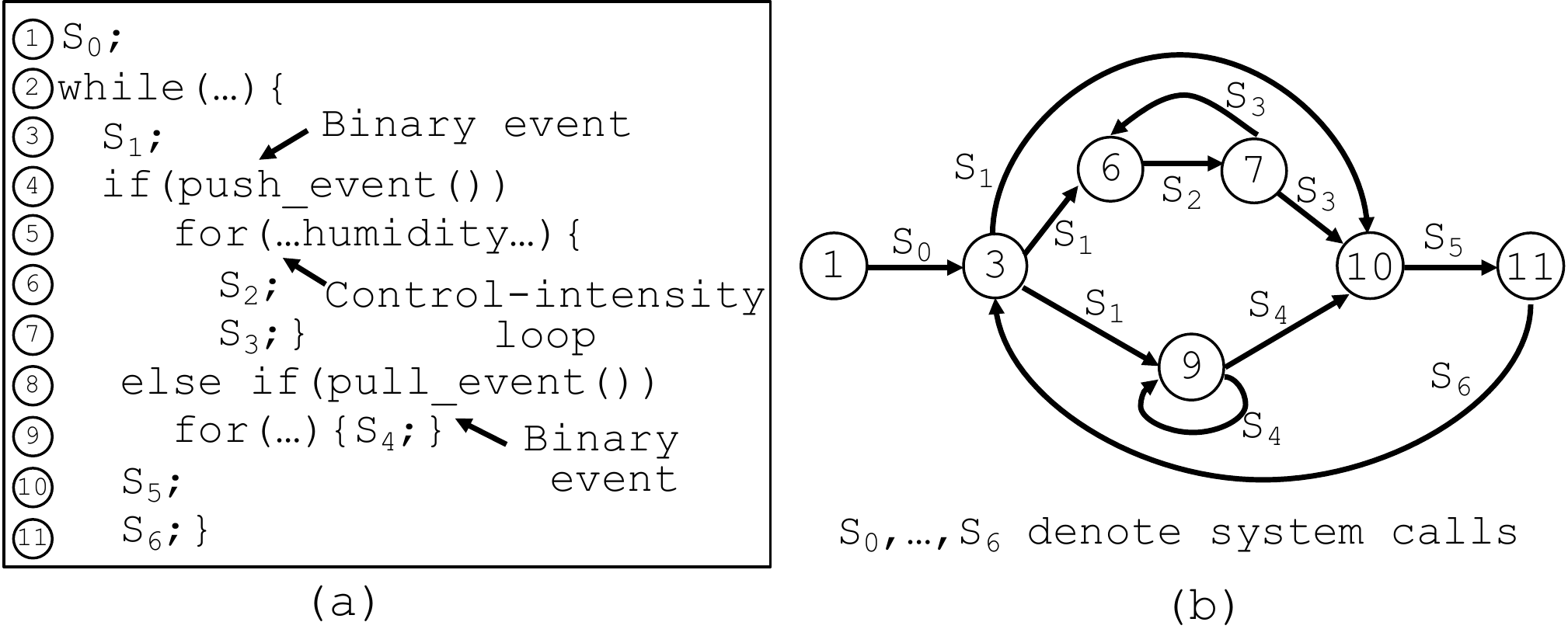} %\vspace{-8pt}
	\caption{\textcolor{clmark}{System-call based finite-state automaton (\FSA{}) model: (a) the example program of Listing~\ref{lst:syringe} at the system-call level; (b) the corresponding \FSA{} model.}} %Line numbers map to nodes in the program's \FSA{}.
	\label{pfsa:fsa}  %\vspace{-8pt}
\end{figure}

In an execution trace, given the $k_{th}$ system call $S_k$ and the PC value $pc_k$ from which $S_k$ was made, the invocation of $S_k$ results in a transition from the previous state $pc_{k-1}$ to $pc_k$ which is labeled with $S_{k-1}$. Fig.~\ref{pfsa:fsa}(a) shows a pictorial example program corresponding to Listing~\ref{lst:syringe} at the system-call level, where system-calls are denoted by $S_0$,\ldots,$S_6$, and states are represented by integers (\ie, line numbers). Suppose we obtain three execution sequences, $\frac{S_0}{1} \frac{S_1}{3} \frac{S_2}{6} \frac{S_3}{7} \frac{S_2}{6} \frac{S_3}{7} \frac{S_5}{10} \frac{S_6}{11}$, $\frac{S_0}{1} \frac{S_1}{3} \frac{S_4}{9}\frac{S_4}{9}\frac{S_5}{10} \frac{S_6}{11}$, and $\frac{S_0}{1} \frac{S_1}{3} \frac{S_5}{10} \frac{S_6}{11} \frac{S_1}{3}\frac{S_5}{10} \frac{S_6}{11}$, the learnt \FSA{} model is shown in Fig.~\ref{pfsa:fsa}(b), where each node represents a state and each arc represents a state transition.

Our \EFSA{} model extends \FSA{} with external context constraints, where event-dependent state transitions in \FSA{} are labeled with event constraints. 
We formally define the \EFSA{} model as a six-tuple: $(S, \Sigma, s_0, F,  E, \delta)$. $S$ is a finite set of states which are PC values, and $\Sigma$ is a finite set of system calls (\ie, input alphabet). 
$s_0$ is an initial state, and $F$ is the set of final states.
$E$ represents a finite set of external events, which can affect the underlying execution of a control program. 
$\delta$ denotes the transition function mapping $S\times\Sigma\times E$ to $S$.
Note that a state transition may come with multiple physical events (referred to as a composite event). Thus, the input alphabet can be expressed as a cartesian product: $E=E_1\times E_2\times\cdots\times E_n$, where the input $E$ consists of $n$ concurrent physical events. In particular, we consider the non-occurrence (not-happening) of one or more events as an implicit event in \EFSA{}.

%where details are introduced in Sec.~\ref{sec:implementation}. 
  
%After discovering statement-level (\ie, instruction-level) event dependence,  . 
%Our event identification module identifies the line numbers in source code where an event is involved. Then, the event dependence analysis outputs the line numbers of event dependent statements. 

\begin{figure}[h] %\vspace{-8pt}
	\centering 
	\includegraphics[width=0.8\columnwidth]{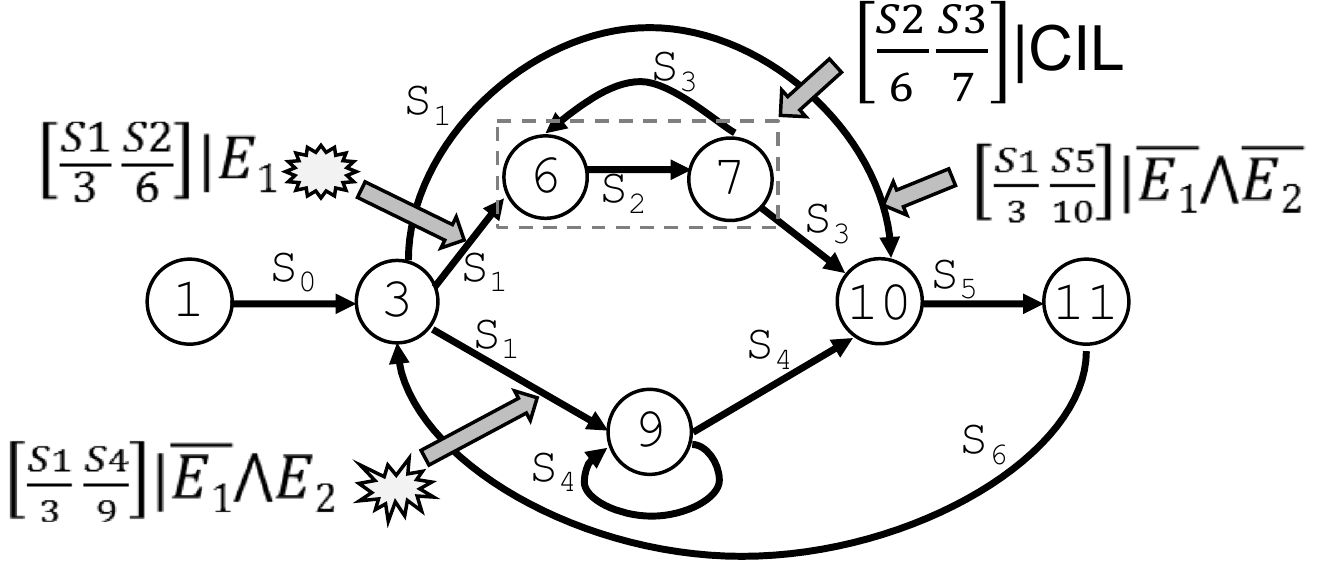} %\vspace{-8pt}
	\caption{\textcolor{clmark}{An example of the \EFSA{} model, where \texttt{$E_1$} represents \texttt{push\_event}, \texttt{$E_2$} represents \texttt{pull\_event}, and \texttt{CIL} represents the control-intensity event/loop.}}
	\label{EventFSA} %\vspace{-6pt}
\end{figure}

Fig.~\ref{EventFSA} shows an example of \EFSA{} model corresponding to the \FSA{} example in Fig.~\ref{pfsa:fsa}, where an event dependent transition is labeled by "$[\frac{System~Call}{PC}]|$Events". 
In this example, there are two binary events and one control-intensity event. Through the event dependence analysis, we identify that lines 5-7 (where $S_2$ and $S_3$ are invoked) and line 9 (where $S_4$ is invoked) are dependent on the binary events \texttt{$E_1$} and \texttt{$E_2$}, respectively. To avoid redundancy, we associate a binary event to the first state transition in \FSA{} that is dependent on it. 
In Fig.~\ref{EventFSA}, we identify binary-event dependent state transitions $[\frac{S_1}{3}\frac{S_2}{6}]|$\texttt{$E_1$}, $[\frac{S_1}{3}\frac{S_4}{9}]|$ \texttt{$\overline{E_1}\wedge E_2$}, and a control-intensity-event dependent control intensity loop $[\frac{S_2}{6}\frac{S_3}{7}]|$\texttt{$CIL$}. It also contains an implicit event dependent transition $[\frac{S_1}{3}\frac{S_5}{10}]|$($\overline{E_1}\wedge\overline{E_2}$).

\subsection{From Event-Annotated CFG to \textit{e}FSA}\label{sec:design:efsa2}

To construct an \EFSA{} model, we need to identify event-dependent state transitions at the system call level in \FSA{}. Towards this end, we apply the event dependence analysis results (described in Sec.~\ref{sec:design:epfsa:event} and \ref{sec:analysis}) to transform instruction-level dependencies in LLVM IR to the state transition dependencies in \FSA{}. Such a mapping might be achieved through static analysis, \eg, passing over the parse tree to search for system call invocations. However, a static analysis based approach requires the modifications of gcc compiler or system call stubs, and even requires hand-crafted modifications for library functions~\cite{Wagner:2001:SP,Lam:2004:RAID}. In \EFSA{}, we adopt a dynamic profiling based approach to discover event dependent state transitions. We first transform instruction-level event dependencies in LLVM IR to statement-level dependencies in source code with line numbers. Then, we map line numbers and file names to return addresses (\eg, by using the \texttt{addr2line} tool) that are collected in the dynamic profiling phase when the \FSA{} model is constructed. In this way, we obtain the system call level event-dependent state transitions in \FSA{}. Subsequently, we augment the event-driven information over the underlying \FSA{}, and finally construct the \EFSA{} model.

\subsection{Security Policies in \textit{e}FSA}
%\subsection{Discovering Event Dependent State Transition}

\EFSA{} expresses causal dependencies between physical events and program control flows. By checking execution semantics (\ie, enforcing cyber-physical security policies) at runtime, \EFSA{} improves the robustness against data-oriented attacks by increasing the difficulties that an attack could bypass the anomaly detection.

For state transitions that are dependent on binary events, the cyber-physical policy enforcement is to make sure the return values of binary events reflects the ground truth sensor measurements.  
For control intensity loops that are dependent on control-intensity events, our approach is based on the control intensity analysis, which models the relationship between the observable information in cyber space (\ie, system-calls) and sensor values in physical space. \EFSA{} then enforces the policy that the observed control intensity should be consistent with the \emph{trend} of sensor value changes.

% we enforce security policies to check whether the \emph{trend} of sensor value changes derived from cyber space is consistent with the sensor measurements in physical space.
 
\subsection{Control Intensity Analysis}
The main challenge for detecting runtime control intensity anomalies lies in that, given system call traces of a control program, we need to map the control intensity to its reflected sensor measurements, where only the number of loop iterations in a control intensity loop is available.
To this end, we first obtain the number of system calls invoked in each loop iteration. Then, we model the relationship between sensor measurements and the amount of system calls in a control intensity loop through a regression analysis.

{\em Execution Window Partitioning and Loop Detection:} Typically, control programs monitor and control physical processes in a continuous manner, where the top-level component of a program is composed of an infinite loop. For instance, an Arduino program~\cite{arduino} normally consists of two functions called \texttt{setup()} and \texttt{loop()}, allowing a program consecutively controls the Arduino board after setting up initial values. We define an \textit{execution window} as one top-level loop iteration in a continuous program, and a \textit{behavior instance} as the program activity within an execution window. The term execution window is equivalent to the \textit{scan cycle} in industrial control domain~\cite{McLaughlin:2014:NDSS}. We partition infinite execution traces into a set of behavior instances based on the execution window. The underlying \FSA{} model helps identify loops since it inherently captures program loop structures. We first identify the starting state in the top-level loop of a \FSA{}. Then, once a top-level loop back edge is detected, a behavior instance is obtained.

{\em Regression Analysis:} 
The purpose of the regression analysis is to quantify the relationship between sensor measurements and system call amount in a control intensity loop.
Given the number of system calls invoked in each loop iteration, one straightforward approach is through manual code analysis. 
In this work, we present an approach for automating this process. During the identification of control-intensity events in Sec.~\ref{sec:design:epfsa:event}, we know what sensor types (\ie, sensor reading APIs) are involved in a control intensity loop. In the training phase, we collect normal program traces together with the corresponding sensor values. Then, we perform a simple regression analysis to estimate the relationship between the system call amount (\ie, outcome) and sensor measurements (\ie, explanatory variables) for each control intensity loop. For example, suppose a control intensity loop is triggered by the change of humidity sensor value (details are in Sec.~\ref{sec:eva:result2}). We observe that an increase of humidity results in more iterations of the control intensity loop, where each loop iteration incurs 3 system calls. Thus, we can reversely derive the changes of physical environment by observing the number of iterations in a control intensity loop.

\section{EFSA-based Detection}\label{sec:design:eventverify}

In this section, we present how an \EFSA{}-based anomaly detector detects anomalies particularly caused by data-oriented attacks, and discuss about the design choices of event verification.

\subsection{Runtime Monitoring and Detection}\label{sec:Monitoring}

\begin{figure}[th]
	\centering \vspace{-8pt}
	\includegraphics[width=0.99\columnwidth]{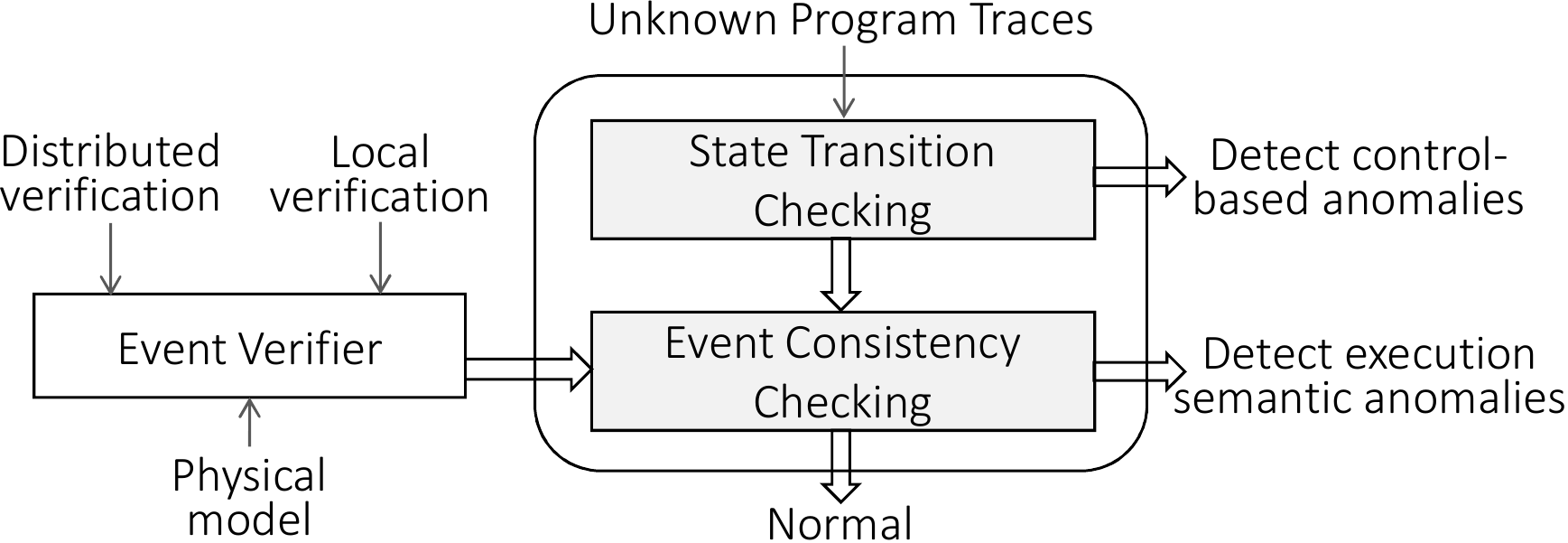}\vspace{-2pt}
	\caption{\EFSA{}-based anomaly detection} \vspace{-8pt}
	\label{fig:testing}
\end{figure} 

Our anomaly detector traces system calls as well as the corresponding PC values during the execution of a control program. 
As shown in Fig.~\ref{fig:testing}, the anomaly detection is composed of an event verifier and two checking steps: i) state transition integrity checking against the basic \FSA{} model, and ii) event consistency checking against the event verification in the \EFSA{}-based anomaly detector, which is our new contribution.
\begin{itemize}[noitemsep, leftmargin=*]
\item 
{\bf Event-independent state transition.}
For each intercepted system call, we check if there exists an outgoing edge labelled with the system call name from the current state in \FSA{}. If not, an anomaly is detected. 
If the current state transition is not event-dependent, we move the current state of the automaton to the new state. This basic state-transition checking has been shown to be effective against common types of control-oriented attacks (\eg, code injection attacks or code-reuse attacks~\cite{Francillon:2008:CIA}) which violate control flow integrity of the model.

\item
{\bf Event-dependent state transition.}
In case of an event dependent state transition according to the \EFSA{} model, we first perform the above basic state-transition checking. More importantly, with the help of the event verification (discussed in Sec. \ref{sec:Verification}), we then check the consistency between the runtime execution semantics and program's behavior, \ie, whether a specific physical event associated with this event-dependent state transition is observed in the physical domain. This step can detect stealthy data-oriented attacks that follow valid state transitions but are incompatible with the physical context. 
Another important aspect is the selection of event checkpoints. To avoid redundant checking, we set the checkpoint for a binary event at its first event-dependent state transition. For a control-intensity event, we perform the event checking after it jumps out of the control intensity loop.
\end{itemize}

\subsection{Event Verification Strategies}\label{sec:Verification}
The objective of event verification is to detect event spoofing caused by runtime data-oriented software exploits.
Event verification is highly application specific, and it is actually orthogonal to the \EFSA{} model itself. We describe several possible approaches for verifying physical context. 
\begin{itemize}[noitemsep, leftmargin=*]
\item {\em Local event verification:} which is able to detect the inconsistency between program runtime behavior and cyber-physical execution semantics. For example, the monitor re-executes a binary-event function to confirm the occurrence of the event. To detect control intensity anomalies, the monitor retrieves sensor measurements and compares them against the derived sensor values from system call traces. There may exist false positives/negatives due to sensor's functional failures in
practice. 
\item
{\em Distributed event verification:} which assesses the physical context by exploiting functionally and spatially redundancy of sensors among co-located embedded devices. Since sensor data normally exhibit spatio-temporal correlation in physical environments, it increases the detection accuracy by involving more event verification sources.

\item
{\em Physical model based verification:} which is complementary to the runtime event verification. Cyber-physical inconsistency may be detected based on physical models~\cite{Urbina:2016:LIS}. For example, one may utilize fluid dynamics and electromagnetics as the basic laws to create prediction models for water system~\cite{Hadziosmanovic:2014} and power grid~\cite{Liu:2009:FDI}. Based on the prediction models and predefined threat constraints, these methods can then check whether the predicted environment values are consistent with a control system's behavior. 
\end{itemize}

\section{Implementation}\label{sec:implementation}
To demonstrate the feasibility of our approach, we have implemented a prototype with around 5K lines in C/C++, Bash and Python codes, including the trace collection and preprocessing, event identification and dependence analysis, \EFSA{} model construction, and runtime anomaly detection modules. Our prototype uses multiple off-the-shelf tools and libraries in Linux.

We choose Raspberry Pi 2 with Sense HAT as the main experimental platform, which is a commonly used platform for building embedded control applications~\cite{McLaughlin:2014:NDSS, Abera:CFLAT:2016, Tan:EMSOFT:2016}. Sense Hat, an add-on board for Raspberry Pi, provides a set of environmental sensors to detect physical events including pressure, temperature, humidity, acceleration, gyroscope, and magnetic filed. During the training phase, we collect program traces on Raspberry Pi and perform the \EFSA{} model construction on a Linux Desktop (Ubuntu 16.04, Intel Xeon processor 3.50GHz and 16GB of RAM). In the monitoring phase, the anomaly detector is deployed on Raspberry Pi to detect runtime control-based or data-oriented attacks. In the following, we present key implementation aspects in our prototype.

%As a special case, we conduct experiments for post-mortem analysis of anomalous behaviors on a commercial drone to demonstrate how \EFSA{} can be applied to network event-triggering scenarios (where details can be found in Sec.~\ref{sec:eva}).

\textbf{Dynamic Tracing.} We use the system tool \texttt{strace-4.13} to intercept system call of a running control program. To obtain the PC value from which a system call was invoked in a program, we need to go back through the call stacks until finding a valid PC along with the corresponding system call. 
We compile \texttt{strace} with \texttt{-libunwind} support, which enables stack unwinding and allows us to print call stacks on every system call.

\textbf{Event Identification and Dependence Analysis.}
Our event identification and dependence analysis tool is implemented within the Low Level Virtual Machine (LLVM)\footnote{http://llvm.org/} compiler infrastructure, based on an open source static slicer\footnote{https://github.com/mchalupa/dg} which builds dependence graph for LLVM bytecode. An advantage of using LLVM-based event dependence analysis is that, our tool is compatible with multiple programming languages since LLVM supports a wide range of languages. Our event identification module identifies the line numbers in source code where an event is involved. Then, the event dependence analysis outputs the line numbers of event dependent statements.

%After discovering statement-level (\ie, instruction-level) event dependence, we next identify event-dependent state transitions (\ie, system call level) in \FSA{}. By using the \texttt{addr2line} tool, we could map line numbers and file names to return addresses (\ie, PC values) that are collected in dynamic profiling phase. Subsequently, we augment the event-driven information over the underlying \FSA{}, and finally construct the \EFSA{} model. 

\iffalse
%[???another alternative is to use gdb ???]
\lstset{language=C,
	keywordstyle=\color{blue},
	stringstyle=\color{red},
	commentstyle=\color{green},
	numbers=left,
	stepnumber=1,    
	firstnumber=1,
	tabsize=1,
	numberfirstline=true
}
\captionsetup{justification=raggedright}
\captionsetup[lstlisting]{position=bottom}
\noindent
\begin{figure}[!h]
	\vspace{-18pt}
	\centering%
	\begin{minipage}{.45\columnwidth} 
		%	\captionsetup{width=0.99\columnwidth} caption=Example  ,   backgroundcolor = \color{magnolia}
\begin{lstlisting}[captionpos=b, basicstyle=\scriptsize, xleftmargin=5em,frame=single,framexleftmargin=2.2em, ]{Name}
int Event(){
	return 1;
}
void handleEvent(){
	return;
}
int main(){   
	if(Event()){
		handleEvent();		
	}
	return 1;
}
\end{lstlisting}
		~~~~~~~~~~~~~~~~~~~~~~~~(a)
	\end{minipage}
	\hfill
	\begin{minipage}{.45\columnwidth}
		\centering
		\resizebox{.77\linewidth}{!}  { \includegraphics[width=\columnwidth]{LLVMExample.pdf}}
		\\(b)
	\end{minipage} 
	\vspace{-8pt}
	\caption{Example of the event dependency analysis}
	\label{fig:llvm}
	\vspace{-10pt}
\end{figure}

\fi

\textbf{Anomaly Detector with Event Verification.} In our prototype, we implement a proof-of-concept near-real-time anomaly detector using named pipes on Raspberry Pi, including  both local and distributed verifications (corroboration with single or multiple external sources). We develop a sensor event library for Raspberry Pi Sense Hat in C code, based on the sensor reading modules in \texttt{experix}\footnote{http://experix.sourceforge.net/} and \texttt{c-sense-hat}\footnote{https://github.com/davebm1/c-sense-hat}. The event library reads pressure and temperature from the LPS25H sensor, and reads relative humidity and temperature from the HTS221 sensor, with maximum sampling rates at 25 per second. Our local event verifier calls the same event functions as in the monitored program, and locally check the consistency of event occurrence. In the distributed event verifier, we deploy three Raspberry Pi devices in an indoor laboratory environment. We develop a remote sensor reading module which enables one device to request realtime sensor data from neighbouring devices via the sockets communication.

\section{Experimental Validation}\label{sec:eva}
%We evaluated  XX %We also discuss the security analysis of \ProjectName{}.

%\textcolor{red}{compare with n-gram (no ret addr), compare tracing overhead with without unwind? }

%\textcolor{red}{new approach of control intensity analysis, How accurate} 

%[??details of x-means clustering, threshold??]
%[??Baseline??]

%We evaluate our prototype implementation from multiple perspectives. In addition, to evaluate the effectiveness of \ProjectName{} on anomaly detection for general-purpose programs, we thoroughly investigate the detection accuracy of \ProjectName{} using synthetic program traces. 

%*** [Long, mention youtube URL, overhead data, ] ***

We conduct CPS case studies, and evaluate \ProjectName{}'s detection capability against runtime data-oriented attacks.
Our experiments aim to answer the following questions:
\begin{itemize}[noitemsep, leftmargin=*]
\item \textcolor{clmark}{What is the runtime performance overhead of \ProjectName{}, including the model training overhead, system-call tracing overhead and detection latency (Sec.~\ref{sec:eva:result3})?}
\item Whether \ProjectName{} is able to detect different data-oriented attacks (Sec.~\ref{sec:eva:result1} and~\ref{sec:eva:result2})? 
%We also provide a video demo to demonstrate \ProjectName{}'s detection capability\footnote{\scriptsize https://youtu.be/-VEjidSgGIc}.
\item \textcolor{clmark}{How feasible is the event-aware n-gram model as an alternative instantiation of the \emph{Orpheus} framework (Sec.~\ref{sec:eva:ngram})?
Whether \ProjectName{} can be generalized to detect network event injection attacks (Sec.~\ref{sec:eva:UAV})?}
  
\end{itemize}

\subsection{CPS Case Studies}\label{sec:eva:casestudey}

\textbf{Solard\footnote{https://github.com/mrpetrov/solarmanpi}.} It is an open source controller for boiler and house heating system that runs on embedded devices. The controller collects data from temperature sensors, and acts on it by controlling relays via GPIO (general purpose input/output) pins on Raspberry Pi. Control decisions are made when to turn on or off of heaters by periodically detecting sensor events. For example, \texttt{CriticalTempsFound()} is a pre-defined binary event in Solard. When the temperature is higher than a specified threshold, the event function returns \verb1True1.
\\
\textbf{SyringePump\footnote{https://github.com/control-flow-attestation/c-flat}.} 
It was developed as an embedded application for Arduino platform. Abera~\etal\cite{Abera:CFLAT:2016} ported it to Raspberry Pi. 
The control program originally takes remote user commands via serial connection, and translates the input values into control signals to the actuator. SyringePump is vulnerable since it accepts and buffers external inputs that might result in buffer overflows~\cite{Abera:CFLAT:2016}.
We modify the syringe pump application, where external inputs are sent from the control center for remote control, and environmental events drive the pump's movement. Specifically, in the event that the relative humidity value is higher than a specified threshold, the syringe pump movement is triggered. In addition, the amount of liquid to be dispensed is linearly proportional to the humidity value subtracted by the threshold. Such sensor-driven syringe pumps are used in many chemical and biological experiments such as liquid absorption measurement experiment.
%\\          
                                                                                                
%\textbf{AR.Drone 2.0 UAV\footnote{\scriptsize https://www.parrot.com/us/drones/parrot-ardrone-20-elite-edition}.} 
%UAVs are typical cyber-physical systems with a tight integration of physical process, computational resource, measurement and communication capabilities. 
%The Parrot AR.Drone 2.0 is a remote controlled quadrocopter, where the control unit receives commands from the remote ground station, monitors and controls the system status to coordinate the flight. Rodday~\etal~\cite{uav:2016:Rodday} exploit security vulnerabilities of the AR.Drone to inject malicious packets and control the Drone. In our experiment, we reproduce the command spoofing attacks to AR.Drone. 

\subsection{Training and Runtime Performance}\label{sec:eva:result3}
In the training phase, we collect execution traces of Solard and SyringePump using training scripts that attempt to simulate possible sensor inputs of the control programs. By checking Solard and SyringePump's source codes, our training scripts cover all execution paths. 
%In the testing phase, we test SyringePump and Solard against two runtime data-oriented attacks. All these attacks do not violate the \FSA{}'s state transition integrity. 

%Since AR.Drone allows a connection to the Telnet port which leads to a root shell, we are able to deploy the \texttt{strace} tool to collect system call traces of the UAV control program. We collect execution traces of AR.Drone by running it using the public testing script\footnote{\scriptsize https://github.com/felixge/node-ar-drone}, which sequentially sends different control commands to the drone. The control program (\ie, \texttt{program.elf}) forks 31 child processes, where we separate system call traces for each process. 

%In AR.Drone, the system call types involved in the process that handles remote commands are quite limited and the program logic is rather simple. Thus, we can easily construct the corresponding \EFSA{} model by taking advantage of network protocol interactions (\ie, network API semantics~\cite{Zhuang:2014:NND}).
%Sample traces of the case studies are included in the accompanying materials for this paper\footnote{\scriptsize Sample system call traces are provided at: https://goo.gl/x6xNbm}.

We first measure the time taken for training models in our prototype, where the main overhead comes from the event dependence analysis. Table~\ref{eva:overhead} illustrates \ProjectName{}'s program analysis overhead in the training phase. For comparison purpose, we deploy the LLVM toolchain and our event dependence analysis tool on both Raspberry Pi and Desktop Computer (Intel Xeon processor \@3.50GHz and 16GB of RAM). From Table~\ref{eva:overhead}, Raspberry Pi takes much longer time (more than 150 times) than desktop computer to complete the program dependence analysis task. It only takes $0.745$s and $0.0035$s for event dependence analysis of Solard (46.3 kb binary size) and SyringePump (17.7 kb binary size) on a desktop computer, respectively. 
Since Solard and SyringePump run in a continuous manner and thus generate infinite raw traces. The model training overhead is measured by how much time it takes for training per MByte raw trace. Results show that it takes less than $0.2$s to process 1 MByte traces on the desktop computer. The number of states in Solard's and SyringePump's \EFSA{} is 34 and 65, respectively (not including system-calls in the initialization before entering the \texttt{main} function).

\begin{table}[!htb]
	%\vspace{-8pt}
	\begin{center}
		\resizebox{8.5cm}{!}{
			\begin{tabular}{ >{\centering\arraybackslash}m{2.0cm} >{\centering\arraybackslash}m{2.8cm}  >{\centering\arraybackslash}m{2.8cm}  }
				\cline{1-3}
				& \multicolumn{2}{c}{Event Dependence Analysis}  \\
				\cline{2-3}
				\cline{2-3}
				\cline{2-3}
				& Desktop Computer & Raspberry Pi 2 \\
				\hline
				\hline        
				Solard & $0.745$s & $109.975$s \\
				\hline
				SyringePump  & $0.0035$s & $1.726$s \\
				\hline
				%\vspace{-16pt}
			\end{tabular}
		}
	\end{center} \vspace{-10pt}	
	\caption{Average delay overhead in training phase}
	\label{eva:overhead}
	%\vspace{-15pt}	
\end{table}

\begin{figure}[th]
 	\centering %\vspace{-8pt}
 	\includegraphics[width=0.58\columnwidth]{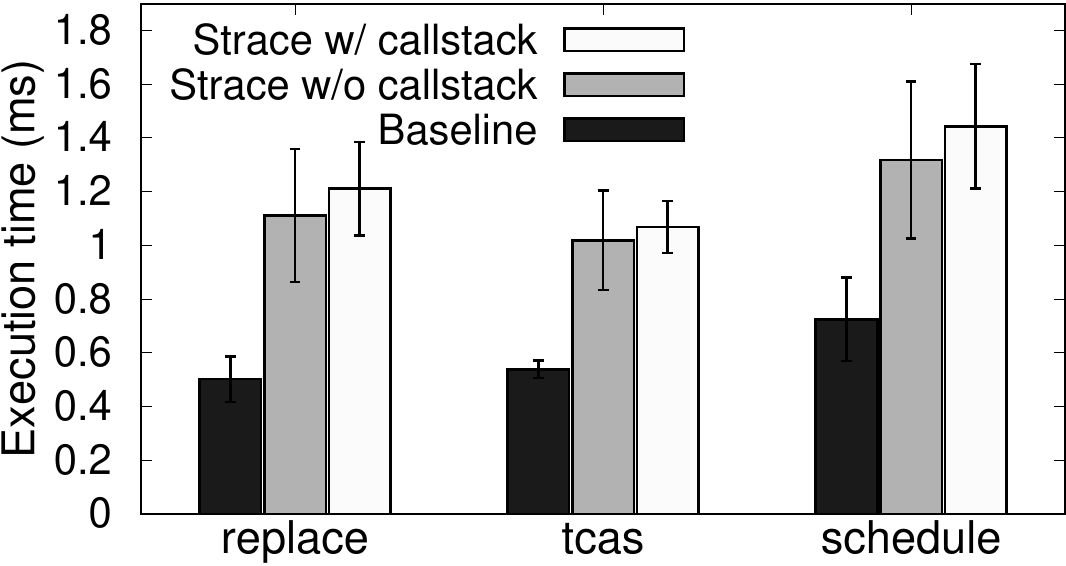}\vspace{-8pt}
 	\caption{\textcolor{clmark}{System-call tracing overhead on Raspberry Pi}} \vspace{-5pt}
 	\label{fig:tracingoverhead}
 \end{figure}

\textcolor{clmark}{
Next, we measure the performance overhead incurred by \ProjectName{}'s anomaly detector on Raspberry Pi, including the system-call tracing overhead and anomaly detection overhead. The system-call tracing overhead has no difference between \FSA{} and \EFSA{}, which incurs $1.5$x$\sim$$2$x overhead in our case studies. To comprehensively measure the runtime system-call tracing overhead, we further experimentally compare the tracing overhead on Raspberry Pi using three utility applications (\ie, \texttt{tcas} (1608 test cases), \texttt{replace} (5472 test cases), and \texttt{schedule} (2650 test cases)) from the Software-artifact Infrastructure Repository (SIR) benchmark suite~\cite{sir}. 
Fig.~\ref{fig:tracingoverhead} shows the results, which measure the elapsed time between the entry and exit points in the three utility applications.
The baseline refers to the execution time without tracing. The runtime performance overhead of \texttt{strace} shows around 96\% slowdown on average. When tracing
the callstack information on every system-call, it yields around 112\% slowdown. We discuss the tracing overhead limitation in Sec.~\ref{sec:limit}.}

\iffalse
\begin{table}[!htb]
	\begin{center}
		\resizebox{7.8cm}{!}{
			\begin{tabular}{ >{\centering\arraybackslash}m{1.5cm} >{\centering\arraybackslash}m{1.3cm}  >{\centering\arraybackslash}m{1.3cm} >{\centering\arraybackslash}m{1.9cm}   }
				\hline
				{Application} & {Baseline} & {strace} & {strace w/ stack}  \\
				\hline
				\hline        
				tcas & 0.538$ms$ & 1.019$ms$ (1.89) & 1.068$ms$ (1.985) \\
				\hline
				replace & 0.502$ms$ & 1.111$ms$ (2.213) & 1.211$ms$ (2.412)   \\
				\hline
				schedule  & 0.725$ms$ & 1.318$ms$ (1.817) & 1.443$ms$ (1.990) \\
				\hline
			\end{tabular}
		}
	\end{center}	\vspace{-8pt}
	\caption{Average tracing overhead on Raspberry Pi}
	\label{tab:eva:overhead:strace}	\vspace{-8pt}
\end{table}
\fi

%We

%replace_inside_delay.log
%mean:0.001111,sdev:0.000248

%replace_inside_delay_stack.log
%mean:0.001211,sdev:0.000174

%schedule_inside_delay.log
%mean:0.001318,sdev:0.000292

%schedule_inside_delay_stack.log
%mean:0.001443,sdev:0.000232

%tcas_inside_delay.log
%mean:0.001019,sdev:0.000186

%tcas_inside_delay_stack.log
%mean:0.001068,sdev:0.000097

Table~\ref{eva:overhead:runtime} reports the runtime anomaly detection latency results. 
The average delay for each state transition (\ie, each intercepted system call) checking out of more than 1000 runs is around $0.0001$s. It takes $0.063$s on average to perform the local event checking. 
The end-to-end latency for the distributed event checking from each co-located device can be broken down into two main parts: i) network communication around $0.042$s, and ii) sensor reading delay around $0.0582$s. In our experiment, we deploy two co-located devices, and thus the total distributed event checking delay is around $0.212$s.   
It is expected that the overhead of distributed event checking is linearly proportional to the number of event verification sources. 

\begin{table}[!htb]
	%\vspace{-5pt}
	\begin{center}
		\resizebox{9.0cm}{!}{
			\begin{tabular}{ >{\centering\arraybackslash}m{3.99cm} >{\centering\arraybackslash}m{2.2cm}  >{\centering\arraybackslash}m{2.5cm}  }
				\hline
				\cellcolor{lightgray}{Delay (Raspberry Pi 2)} & Mean & Standard Deviation \\
				\hline
				\hline
				FSA State Transition Checking  & $0.00013293$s & $0.00004684$s \\
				\hline        
				Local Event Verification & $0.06279120$s & $0.00236999$s \\
				\hline
				Distributed Event Verification & $0.21152867$s & $0.03828739$s \\
				\hline
				%\vspace{-16pt}
			\end{tabular}
		}
	\end{center} \vspace{-10pt}
	\caption{Runtime detection overhead in the monitoring phase}
	\label{eva:overhead:runtime}
	
\end{table}

\vspace{-22pt}

\subsection{Detecting Attacks on Control Branch}\label{sec:eva:result1} 
In this experiment, we evaluate \ProjectName{}'s security guarantees against control branch attacks. 

\subsubsection{Solard}

In Solard, we engineer a buffer overflow vulnerability and manipulate the temperature sensor values to maliciously prevent the heater from being turned off. This cyber-physical attack is similar to the recent real-world German steel mill attack~\cite{GermanSteelMill}, which may result in a blast furnace explosion.
In this experiment, we attach the Raspberry Pi on an electric kettle (\ie, 1-Liter water boiler). 
The control program keeps monitoring temperature values. When the temperature is lower than 50$^{\circ}C$, it turns on the heater. And when the temperature is higher than 60$^{\circ}C$, where \texttt{CriticalTempsFound()} is supposed to return \verb1True1, it turns off the heater. 
In the monitoring phase, when we detect an event-dependent state transition in \EFSA{} model, the local event verifier performs event consistency checking. 

%(Details of SyringePump case study against control branch attacks can be found in our video demo)

\iffalse
\begin{figure}[ht]
	\centering
	%\vspace{-5pt}
	\subfloat[\small Sensor values in Solard]{
		\includegraphics[width=0.38\columnwidth]{eventmonitor0.pdf}
		\label{fig:eva:eventmonitor1}
	}
	%\hfill
	\subfloat[\small Sensor values in event verifier]{
		\includegraphics[width=0.4\columnwidth]{eventmonitor1.pdf}	
		\label{fig:eva:eventmonitor2}
	}
	\vspace{-8pt}
	\caption{An instance of Solard experiment}
	\label{fig:eva:eventmonitor22}
%	\vspace{-22pt}
\end{figure}
\fi

\begin{figure}[th]
	\centering %\vspace{-8pt}
	\includegraphics[width=0.7\columnwidth]{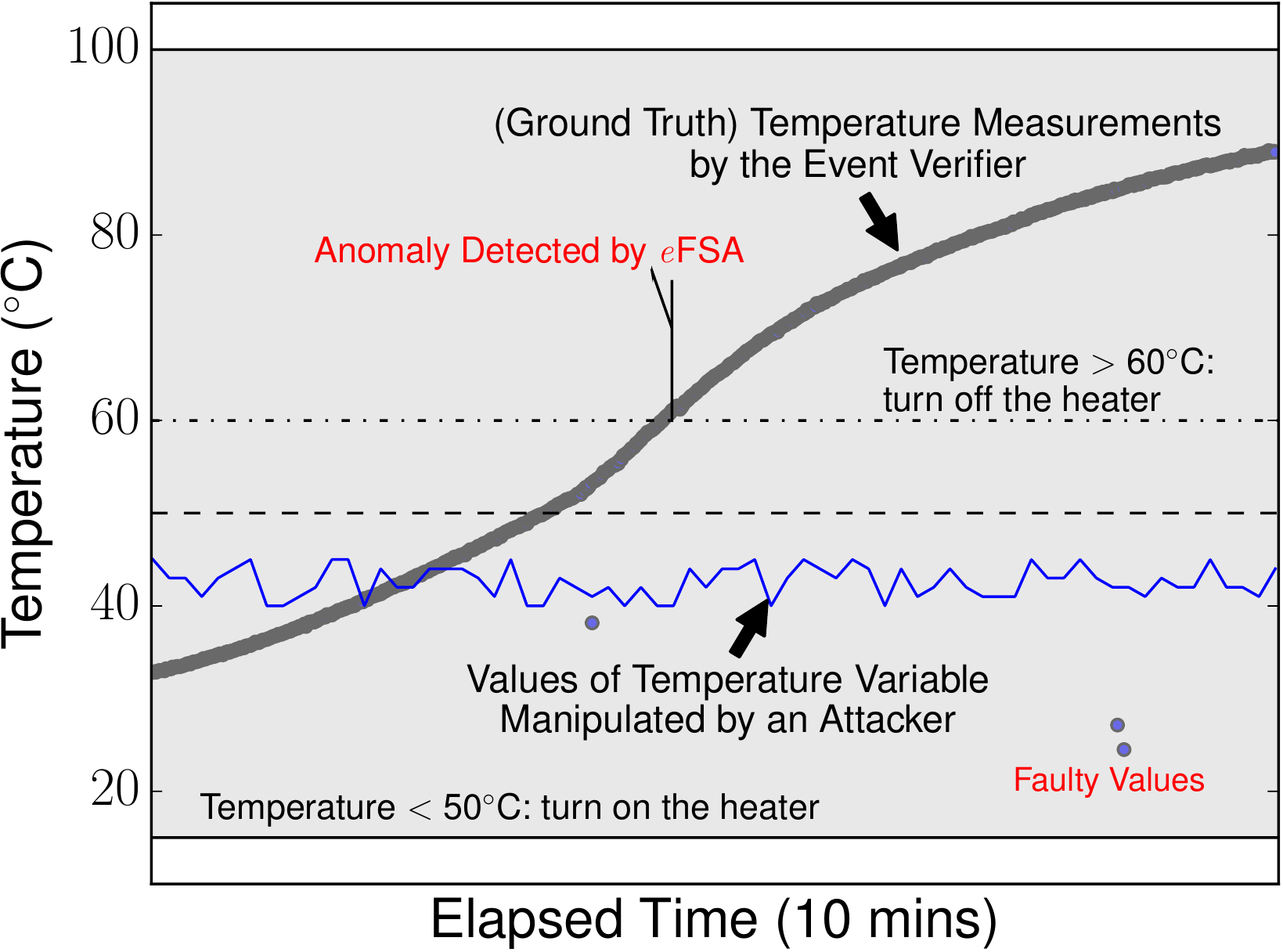}\vspace{-8pt}
	\caption{An instance of Solard experiment} \vspace{-5pt}
	\label{fig:eva:eventmonitor}
\end{figure} 

Fig.~\ref{fig:eva:eventmonitor} illustrates an instance of the Solard experiment. 
We corrupt the temperature sensor values in the range of 40$\sim$45$^{\circ}C$, which falsifies the return value of \texttt{CriticalTempsFound()} to be always \verb1False1. 
In every scan cycle, \ProjectName{} observes a state transition dependent on the not-happening of \texttt{CriticalTempsFound()} (\ie, an implicit event), and thus the event verifier checks the instantaneous temperature value. In our experiment, because the Raspberry Pi does not physically interact with the electric kettle, the ground truth temperature keeps increasing up to more than 80$^{\circ}C$ in Fig.~\ref{fig:eva:eventmonitor}. However, \ProjectName{} successfully raises an alarm at the first moment when it finds a mismatch between the execution semantics (temperature exceeding 60$^{\circ}C$) and program behavior.

We did encounter sensor measurement failures, \eg, isolated dots as shown in Fig.~\ref{fig:eva:eventmonitor}. 
On average, the false sensor measurement rate is lower than 1\% in our experiments. This means that the detection rate and false positive/negative rate would depend on sensors' functional reliability in practice. Existing methods, such as data fusion~\cite{fusion:2009} can be applied to enhance the detection accuracy. 

\subsubsection{SyringePump}
In SyringePump, we set the threshold to 40$rH$, \ie, when the relative humidity value is higher than 40$rH$, it drives the movement of syringe pump by sending control signals to dispense liquid. The buffer overflow attack manipulates the humidity sensor values to purposely trigger \texttt{event-push} control actions without receiving an external event or environmental trigger. Such an attack leads to unintended but valid control flows.

\begin{figure}[th]
	\centering \vspace{-3pt}
	\includegraphics[width=0.9\columnwidth]{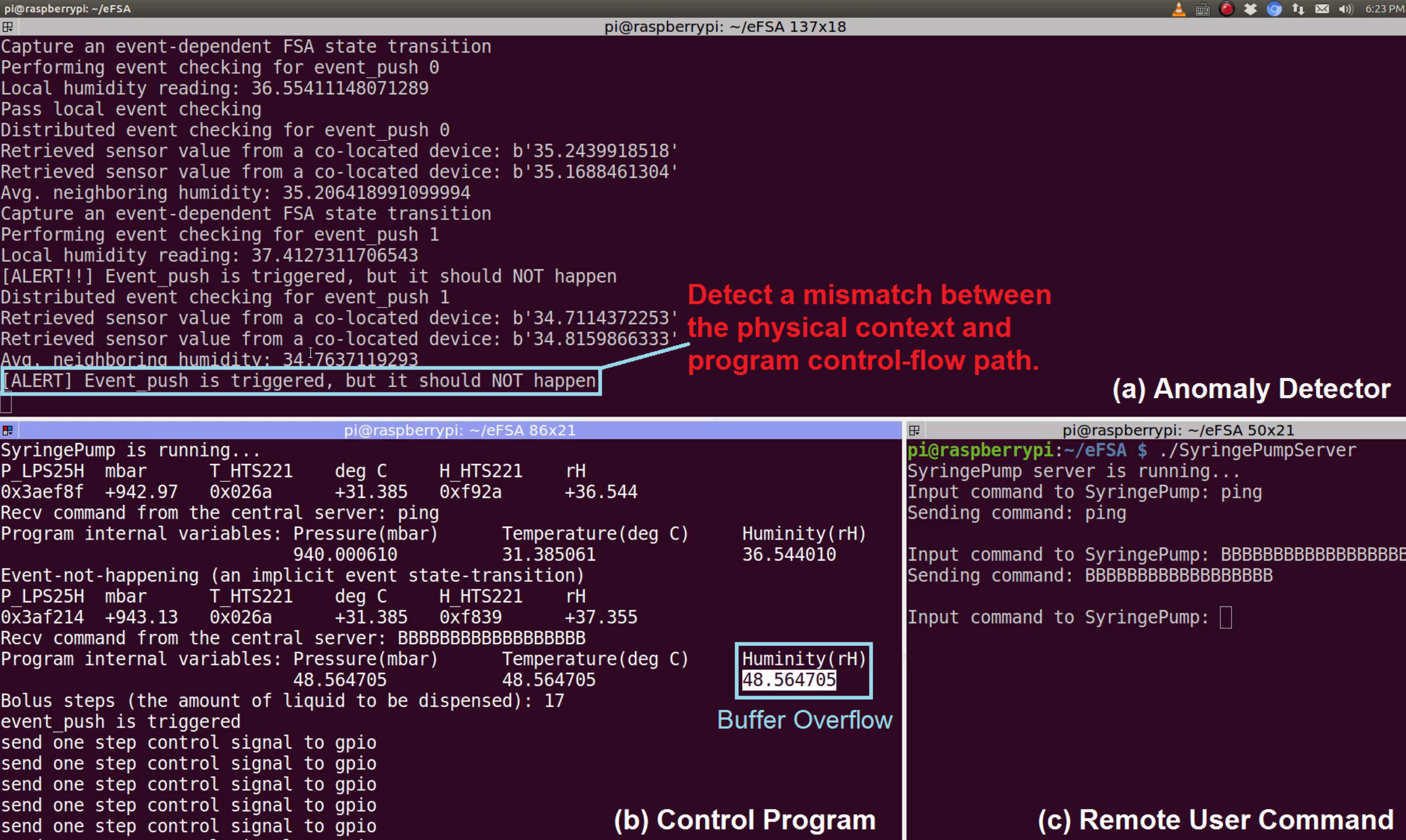}\vspace{-5pt}
	\caption{An instance of SyringePump experiment} \vspace{-3pt}
	\label{fig:eva:examplePump}
\end{figure}

Fig.~\ref{fig:eva:examplePump} illustrates an example of the experiment. 
The remote user command corrupts the humidity sensor value to be 48.56$rH$, which falsifies the return value of \texttt{event-push} to be \verb1True1.  
In case of any event-driven state transition according to \EFSA{}, the event verifier checks consistency between the runtime execution semantics (\eg, the instantaneous humidity value) and program internal state. As shown in Fig.~\ref{fig:eva:eventmonitor}, \ProjectName{} raises an alarm when it finds a mismatch between the execution semantics and program behavior.

\subsection{Detecting Attacks on Control Intensity}\label{sec:eva:result2}

\begin{figure}[th]
	\centering \vspace{-3pt}
	\includegraphics[width=0.8\columnwidth]{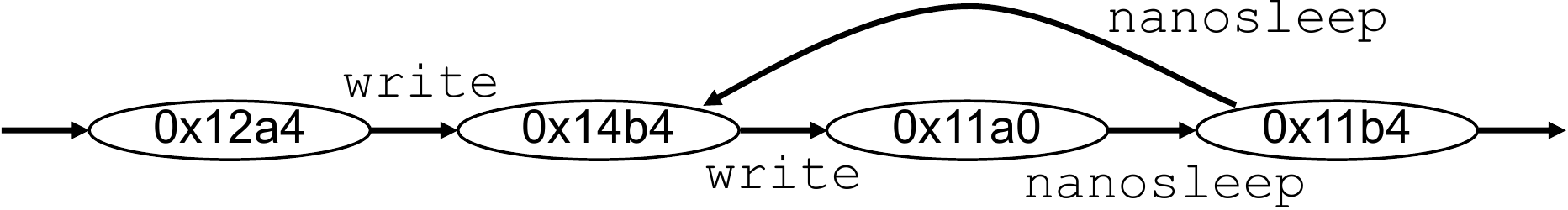}\vspace{-5pt}
	\caption{\textcolor{clmark}{Control intensity loop in \EFSA{} of SyringePump}} \vspace{-3pt}
	\label{fig:eva:controlloop}
\end{figure}

In this experiment, we demonstrate that \EFSA{} is able to detect control intensity attacks with only system call traces. In SyringePump, we set the threshold that triggers the movement of syringe pump to be 30$rH$. The corrupted humidity value determines the amount of liquid to be dispensed, which equals to the humidity value subtracted by 30$rH$ in this test. In the training stage, we obtain the number of system calls invoked in each loop iteration. Then, we model the relationship between sensor measurements and the amount of system calls in a control intensity loop. Through control intensity analysis, we know the number of system calls with no event occurrence is 40 per scan cycle, and each loop iteration (\ie, dispensing a unit of liquid) in the control intensity loop corresponds to 3 system-calls \texttt{write}-\texttt{nanosleep}-\texttt{nanosleep}, as shown in Fig.~\ref{fig:eva:controlloop}. 

%Our attack corrupts the humidity sensor value based on a buffer overflow vulnerability~\cite{Abera:CFLAT:2016} and thus could drive the movement of syringe pump without receiving an external event or environmental trigger (\ie, regardless the relative humidity value is higher than 30$rH$). Meanwhile, the corrupted humidity value also determines the amount of liquid to be dispensed, which equals to the humidity value subtracted by 30$rH$. In the training phase, through control intensity analysis, we know the number of system calls with no event occurrence is 40 per scan cycle, and each loop iteration (\ie, dispensing a unit of liquid) in the control intensity loop corresponds to 3 system calls. 

\begin{figure}[ht]
	\centering
	\vspace{-12pt}
	\subfloat[Humidity and system call traces]{
		\includegraphics[width=0.48\columnwidth]{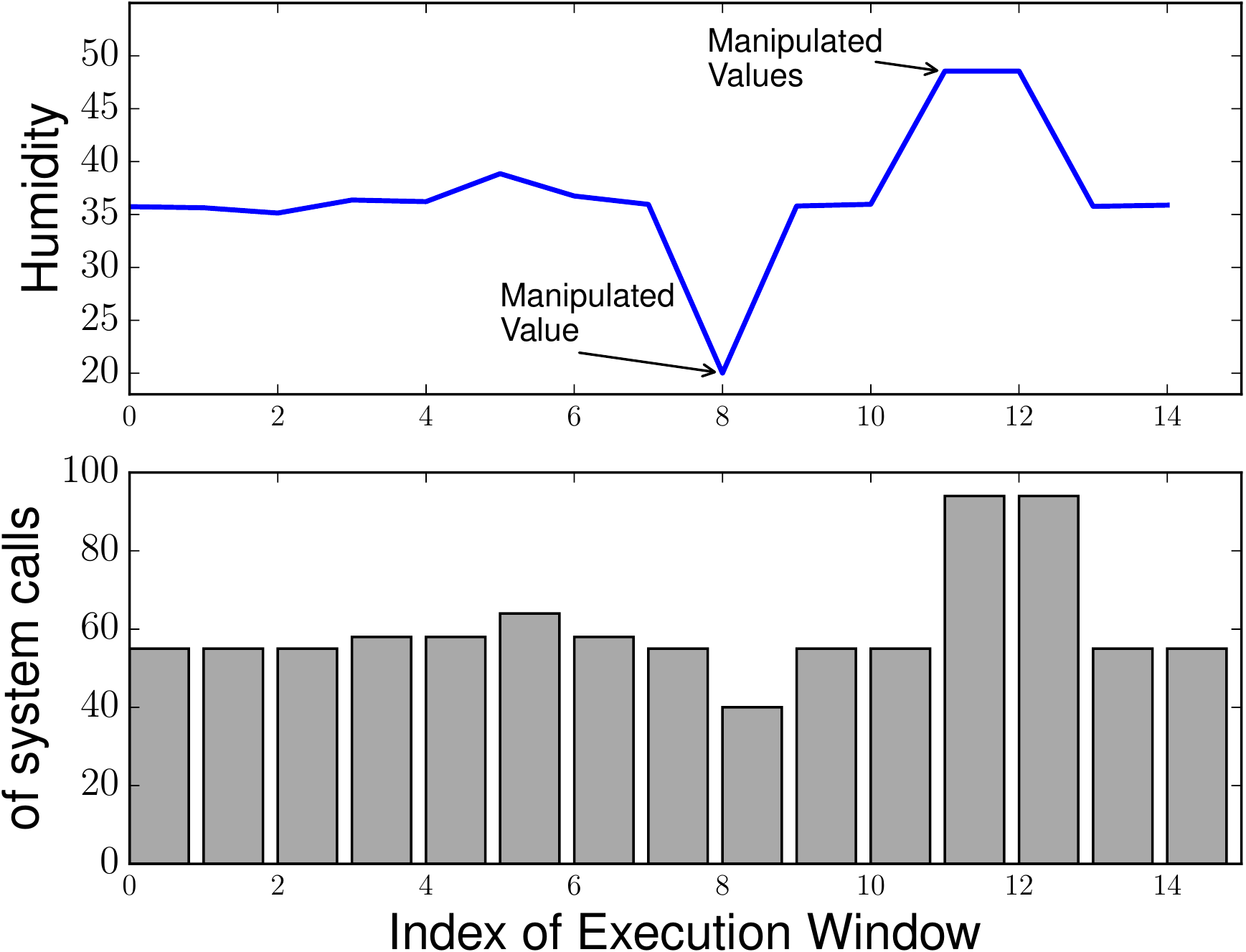}
		\label{fig:eva:syringe1}
	}
	%\hfill
	\subfloat[\EFSA{}'s detection]{
		\includegraphics[width=0.48\columnwidth]{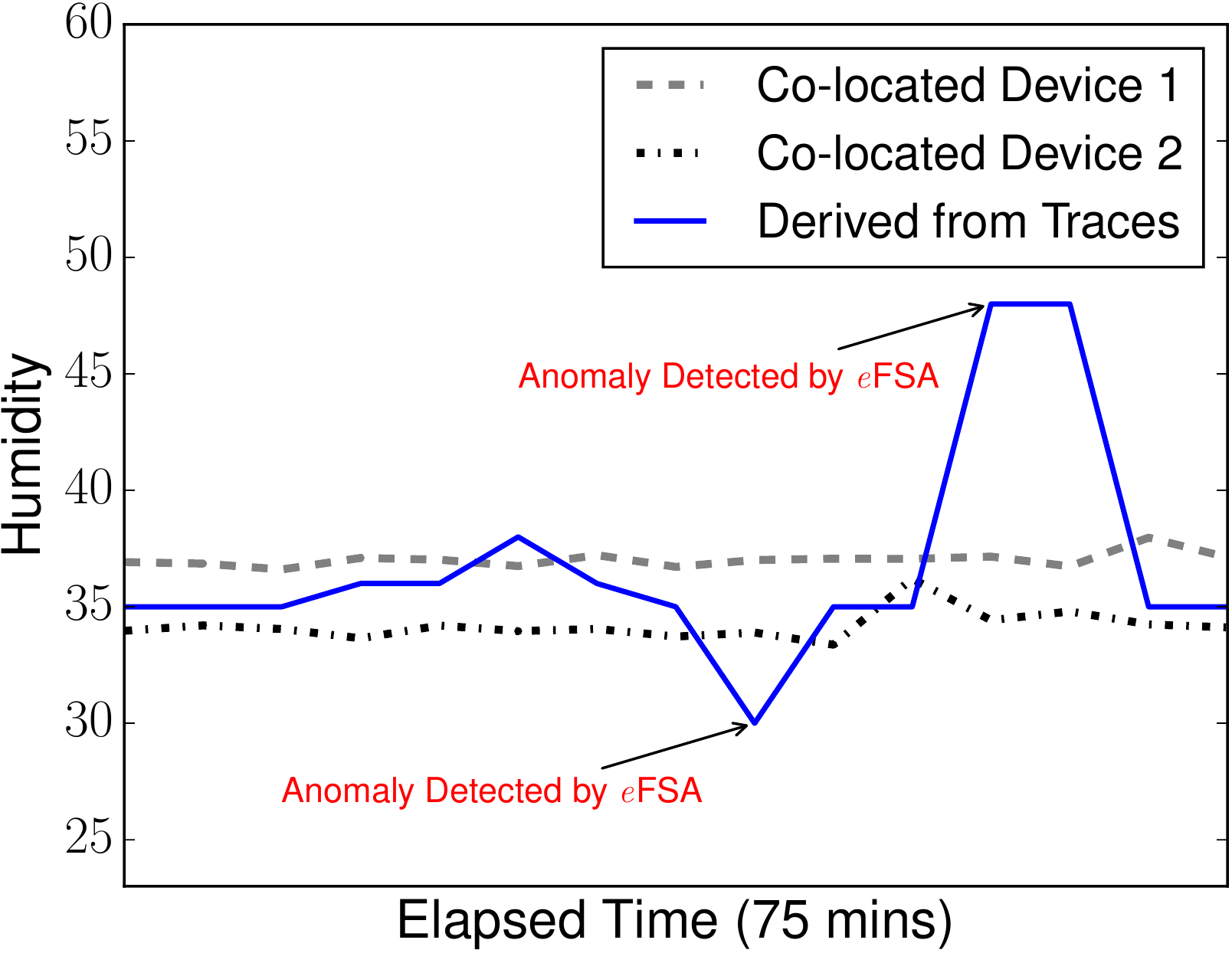}	
		\label{fig:eva:syringe2}
	}
	\vspace{-6pt}
	\caption{An instance of SyringePump experiment with a sampling rate of 5 minutes}
	\label{fig:eva:syringe}
	\vspace{-9pt}
\end{figure}

Fig.~\ref{fig:eva:syringe}(a) shows the value changes of the humidity variable and system call amount per scan cycle of SyringePump. The normal humidity value fluctuates between 34 $rH$ and 38$rH$. As a result, the amount of liquid to be dispensed is subsequently changed, which is reflected by the number of system calls in each control loop. 
We manipulate the humidity values to be 20$rH$ and 48$rH$, respectively. In the monitoring phase, by observing the number of system calls in each control loop, we can reversely derive the changes of physical environment based on our control intensity regression model as shown in Fig.~\ref{fig:eva:syringe}(b). In this test, if the difference between the derived value and the sampled average value from event verifier is larger than 3$rH$, we consider it an anomaly. By checking the humidity measurements from two co-located devices (\ie, denoted as devices 1 and 2), our distributed event verifier detects that the program's runtime behaviors are incompatible with physical contexts. Thus, \EFSA{} successfully detects the control intensity attacks. 

\subsection{Event-Aware N-gram Model}\label{sec:eva:ngram}
\textcolor{clmark}{
To show the feasibility of augmenting the n-gram model with event awareness, we conduct the case study of the event-aware n-gram model, which is an alternative instantiation of the \emph{Orpheus} framework. 
Given the execution traces of SyringePump in the training phase, we construct the n-gram model of system-call sequences (n$\in$[2,$\cdots$,10]). 
For the 2-gram model, there are 35 different 2-grams. \texttt{write}-\texttt{write} is an event-dependent 2-gram. 
However, 15 non-event-dependent \texttt{write}-\texttt{write} system-call sequences are also observed during the monitoring phase, which makes the event checking no longer effective due to the ambiguity. This is because, without the PC (return address) information associated with each system-call, we cannot differentiate the invocation of the same system-calls from different places in the program. As a result, we need to increase the length of the n-gram until the event-dependent n-gram is unique in the model. But there is no guarantee we can always find the unique event-dependent n-gram with increasing the value of $n$. 
In SyringePump, we find a unique event-dependent 4-gram \texttt{write}-\texttt{write}-\texttt{nanosleep}-\texttt{nanosleep}, where the model contains 60 different 4-grams. During the monitoring phase, our anomaly detector performs the event checking whenever this event-dependent 4-gram is observed. 
Nevertheless, with the increase of $n$, the size of the n-gram model also increases. For example, when we increase $n$ to 10, the size of the n-gram model is increased to 96, which is larger than the size of \EFSA{}. From this case study, we demonstrate that it is desirable to use the program counter (PC) information in the basic program behavior model, which significantly increases the resistance against control-flow attacks, but also resolves ambiguities in our event-aware anomaly detection.}

\subsection{Extension of \textit{e}FSA}\label{sec:eva:UAV}
%\subsection{Generalization of Events and Detecting Command Injection Attacks}\label{sec:eva:uav}

\textcolor{clmark}{
Control programs running on embedded devices may receive network events from the control center, and then execute actuation tasks. Though \EFSA{} mainly detects software-exploit based physical event spoofing, it is also applicable to network event-triggering scenarios. 
For example, a ground control station (GCS) sends a command to control an unmanned aerial vehicle (UAV) to change its flight mode. In this case, we consider each type of control command as a specific event, and the \EFSA{} model of a UAV program is augmented with both physical sensing and network command events. Such an \EFSA{} model can detect false command injection attacks against the UAV by checking the consistency of system call traces at a UAV and its GCS, ensuring their system call invocations conforming to the network API semantics~\cite{Zhuang:2014:NND}.}

\textcolor{clmark}{
To demonstrate the applicability of \EFSA{} for detecting network event inconsistency in CPS, we conduct an experiment on AR.Drone\footnote{\scriptsize https://www.parrot.com/us/drones/parrot-ardrone-20-elite-edition}, which is a remote controlled quadrocopter. 
Different from the Solard and SyringePump case studies running on Raspberry Pi, AR.Drone runs on a customized mainboard with embedded Linux.
AR.Drone allows a connection to the Telnet port which leads to a root shell, and thus we are able to deploy the \texttt{strace} tool to collect system call traces of the UAV control program.
In AR.Drone, the system call types involved in the process that handles remote commands are quite limited and the program logic is rather simple. Thus, we can easily construct the corresponding \EFSA{} model by taking advantage of network protocol interactions without program dependency analysis.
Sample traces are included in the accompanying materials for this paper\footnote{\scriptsize Sample system call traces are provided at goo.gl/Wkrdzz}.}
 
%Rodday~\etal~\cite{uav:2016:Rodday} exploit security vulnerabilities of the AR.Drone to inject malicious packets and control the UAV. 
\textcolor{clmark}{
Since AR.Drone runs with an open Wi-Fi, and we know the communication protocol between UAV and the GCS~\cite{uav:2016:Rodday}, we can easily launch a false command injection attack, which could be used to send malicious operational data such as status or control command. We use a laptop (as an attacker) to send fake control commands to the AR.Drone. Meanwhile, we monitor the network traffic at GCS using the Wireshark tool. The event verifier could find inconsistencies between \EFSA{}'s state transitions and network events captured at GCS, and thus detect this type of attacks. In this experiment, we do not intend to use \EFSA{} to raise an alarm at the time of intrusion, instead we aim at detecting the anomalous behavior as a post-mortem analysis.}

%In our experiment, we reproduce the command injection attacks to AR.Drone.

%\subsection{Security Considerations}\label{sec:eva:discussion} ???

\section{Limitation}\label{sec:limit}

Although our work is focused on providing new security capabilities in control-program anomaly detection against data-oriented attacks, in this section, we examine the limitations of our implementation and discuss how our method can be deployed in the near future.

\textbf{Bare-metal CPS Devices}: 
Our anomaly detection system works on the granularity of system calls and it leverages dynamic tracing facilities such as the \texttt{strace} tool, which requires the operating system support. 
An important reason behind our choice is that, the new generation of embedded control devices on the market are increasingly coming with operating systems~\cite{Schwartz:SP:2014, Tan:EMSOFT:2016}. For example, Raspberry Pi devices with embedded Linux OS have been used as field devices in many CPS/IoT applications~\cite{Pi4ICS}. 
Linux-based PLCs for industrial control have emerged to replace traditional PLCs~\cite{REXControl} for deterministic logic execution. 
However, embedded devices may still operate in bare-metal mode~\cite{Abera:CFLAT:2016}, where we can not utilize existing tracing facilities to collect system call traces. 
For traditional PLCs, our security checking can be added to the program logic. 
We can also apply the event checking idea to an anomaly detection system at the level of instructions. We may instrument the original control program with event checking hooks by rewriting its binary, \eg, inserting hooks at the entry of event-triggered basic blocks. We consider it as the future work to extend our design paradigm for fine-grained anomaly detection with binary instrumentation.

\textbf{Tracing Overhead and Time Constraints}: Though system call traces are a common type of audit data in anomaly detection systems, we would like to point out that the conventional software-level system call tracing incurs unnegligible performance overhead to the monitored process~\cite{Feng:SP:2004}. It holds for time-insensitive embedded control applications, \eg, smart home automation, but would be a technical challenge for time-sensitive applications. 
While we employ the user-space \texttt{strace} software to collect system calls in our prototype, tracing tools are orthogonal to our detection design. For performance consideration, alternative tracing techniques may be adopted in replacing \texttt{strace} to improve the tracing performance~\cite{Shu:2015:USP}. 
%Existing work~\cite{Pohlack06towardsruntime} demonstrates that runtime monitoring of hard real-time systems is feasible, where functionality and monitoring requirements are known beforehand and predictable. 
For example, it is possible to improve the performance for system call interposition by modifying the kernel at the cost of increased deployment effort. With the recently unveiled Intel's Processor Trace (PT) and ARM's CoreSight techniques, hardware tracing infrastructures are increasingly embedded in modern processors, which can achieve less than 5\% performance overhead~\cite{CPE:CPE4069}. The recent work, Ninja~\cite{ninja:usenixsecurity17}, offers a fast hardware-assisted tracing on ARM platforms. The overhead of instruction tracing and system call tracing are negligibly small. Therefore, we anticipate that future tracing overhead will be significantly reduced as the hardware-assisted tracing techniques are increasingly used.

\textbf{Lack of CPS Benchmark Programs}: \textcolor{clmark}{Lack of CPS benchmark programs is one of the challenges in CPS security research to perform sound evaluation. 
It is partly because of the diversity of CPS hardware platforms, and the hardware-dependent nature of CPS programs. In addition, safety-critical CPS programs are rarely open-source. 
As a result, existing CPS security research work mainly conducted limited case studies~\cite{Abera:CFLAT:2016}\cite{OEI:Sun:2018}\cite{Munir:TDSC:2018}, instead of a large scale experimental evaluation. We therefore leave the comprehensive evaluation of our approach for future work as more CPS benchmark programs are available. We also would like to exploit the symbolic execution in our control intensity analysis, which may statically derive the relationship between sensor values and the number of control loop iterations. Symbolic execution and fuzzing techniques are also useful to increase code-coverage in the training phase for collecting traces of normal program executions.}

% The research community needs to provide mechanisms to incentivize data sharing and benchmark preparation effort.

%\textcolor{red}{  ours is small scale, for large program [talk about static our analysis, and symbolic execution can help to generate test cases for coverage , symbolic execution help for the control intensity analysis, only based on the pure static analysis]
%	fuzzing and symbolic execution can be used to increase code-coverage for an automated test-case generation. Fuzzing executes the app with random input data, and symbolic execution uses symbolic inputs to perform path-based exploration}

\textbf{Limitation of Detection Capability}: 
\textcolor{clmark}{
From our case studies, we demonstrate that enforcing cyber-physical execution semantics in CPS program anomaly detection is effective to detect both types of data-oriented attacks. The necessary condition is that the observed program behavior at system-call level is incompatible with the current physical context. Simply corrupting non-control data in a program's memory space may be undetectable if the attack does not result in inconsistencies between the physical context and program execution, which is out of the scope in this work. 
Another limitation of our design is that we only detect program anomalies at the system-call level. 
CPS devices may send control signals by directly
writing registers without issuing any system-call, rendering the system-call based detection not working. To overcome this limitation, the \emph{Orpheus} design paradigm can be extended to the instruction level for fine-grained anomaly detection.}
%limitation: can not detect memory corruption data oriented attack but does not incur inconsistency  model sensitive of code change, need to retrain the model

%We use a laptop (as an attacker) to send fake control commands to the AR.Drone. Meanwhile, we collect system calls on both Drone and GCS. The event verifier could find inconsistencies between \EFSA{}'s state transitions and network events (\eg, \texttt{sendto} or \texttt{recvfrom}) retrieved from GCS, and thus detect this type of attacks. In this experiment, we do not intend to use \EFSA{} to raise an alarm at the time of intrusion, instead we aim at detecting the anomalous behavior as a post-mortem analysis.

\textbf{Anomaly Detection as a Service}: 
Embedded devices are resource-constrained compared with general-purpose computers. To reduce detection overhead, the anomaly detection may be performed at a remote server. 
%In this case, a secure daemon running on the CPS device collects runtime event information and system call traces together with PC values, and then sends them to the server for detection. 
We envision deployment involving partnerships between hardware vendors and security service providers (similar to ZingBox IoT Guardian~\cite{ZingBox}), where the security provider is given access to embedded platforms and helps clients to diagnose/confirm violations.
The client-server architecture resonates with the remote attestation in embedded systems, which detects whether a controller is behaving as expected~\cite{Valente:2014:CSA, Abera:CFLAT:2016}. For detection overhead reduction, the remote server may choose when and how frequently to send assessment requests to a control program for anomaly detection. It is also possible to selectively verify a subset of events for the scalability purpose, \eg, only safety-critical events specified by developers are involved.
While the event verifier implementation is not completely automated, our event identification and dependence analysis tool does automate a large portion of event code extraction and eases the developer's burden. We leave automatically generating event verification functions for the anomaly detector as an important part of our future work.

\section{Related Work}\label{sec:relatedwork}

%Our contribution in this work lies at the intersection of two research areas: CPS anomaly detection and program behavior modeling.
%In this section, we briefly summarize related works in these two research areas.
%\vspace{-3pt}
 
%\subsection{Anomaly Detection in CPS}
 
%Our work focuses on anomaly detection for embedded control programs with cyber-physical interactions, \eg, in CPS/IoT applications.

%Since such embedded control system is an integration of cyber and physical components, previous anomaly detection works can be divided into two lines of research: based on i)  behavior models of physical processes; or ii) behavior models of cyber programs/systems. 

Due to the diversity of CPS applications, existing anomaly detection solutions are proposed to detect specific attacks for specific applications, such as smart infrastructures~\cite{Sridhar:2012:PIEEE}, unmanned aerial vehicles~\cite{Mitchell:2014:UAV}, medical devices~\cite{Mitchell:2015:TDSC}, automotive~\cite{Cho:2015:CAC, Rouf:2010:SPV}, industrial control process~\cite{Cardenas:2011:AAP, McLaughlin:2014:NDSS, Urbina:2016:LIS}. The majority of research efforts in this area thus far have concentrated on behavior model-based anomaly detection~\cite{Urbina:2016:LIS}, and can be generally classified into two categories: {\em 1)} cyber model (\eg, program behavior model, network traffic analysis, or timing analysis); {\em 2)} physical model (\eg, range-based model or physical laws). Our proposed \EFSA{} analyzes both the cyber and physical properties of CPS, as well as their interactions. Thus, we refer to it as the cyber-physical model. Table~\ref{relatedwork::detectionmethods} compares representative CPS anomaly detection solutions.

\begin{table*}[!htb]
	\newcommand{\tabincell}[2]{\begin{tabular}{@{}#1@{}}#2\end{tabular}}
	\begin{center}
		\resizebox{17.5cm}{!}{
			\begin{tabular}{  >{\centering\arraybackslash}m{3.3cm} >{\centering\arraybackslash}m{3.0cm} >{\centering\arraybackslash}m{4.5cm} >{\centering\arraybackslash}m{6.0cm} >{\centering\arraybackslash}m{4.0cm}  }
				\hline
				Research Work & Category & Approach & Security Guarantee  & Validation   \\ %Weaknesses Strengths
				\hline
				\hline
				
				Yoon~\etal\cite{CoRR:YoonMCCS15} & Program behavior model (cyber)  & Syscall frequencies  & Frequency-based program control flow anomaly & Raspberry Pi testbed\\
				\hline
				Feng~\etal\cite{Feng:2017:DSN} & Network traffic analysis (cyber)   & Machine learning based traffic analysis & Traffic alteration & Traffic data from a gas pipeline system \\
				\hline
				Zimmer~\etal\cite{Zimmer:2010:TID} & Timing analysis model (cyber) & Static/dynamic timing analysis  & Code injection attacks  & Simulation/Testbed\\
				\hline   
				C-FLAT~\cite{Abera:CFLAT:2016} & Program behavior model (cyber) & Program analysis and instrumentation & Control-oriented attacks and limited non-control-data attacks & Raspberry Pi testbed\\
				\hline 
				
				FT-RMT~\cite{Munir:TDSC:2018} & Redundant execution analysis (cyber) &  Redundant controller and computation & CPS faults or attacks & Simulate steer-by-wire application on embedded Linux \\
 				 
				\hline   
				Hadziosmanovic~\etal \cite{Hadziosmanovic:2014}  &  Range-based model (physical) & Attribute values extracted from network traffic & False data injection attacks & Traffic data from water treatment plants \\
				\hline
				NoisePrint~\cite{Ahmed:2018:NAD} & Noise fingerprinting (physical) & Pattern recognition in sensor and process noise dynamics & Sensor spoofing attacks  &  Two real-world CPS testbeds\\
					\hline
				Cardenas~\etal \cite{Cardenas:2011:AAP} & Physical laws & Linear model derived from training data  & False data injection attacks  & Simulation\\
				\hline
				SRID~\cite{Wang:2014:ESORICS} & Physical laws  & Correlation analysis of system variables.  & False data injection attacks  & Simulation\\
				\hline
				C$^2$\cite{McLaughlin:2013:CSP} & Control policies (physical) &  User specified control policies& Control signal violation & Raspberry Pi testbed  \\
				\hline
				\EFSA{} (Our work) & Cyber-physical model & Event-aware FSA & Data-oriented attacks & Raspberry Pi testbed  \\
				\hline
				\hline
			\end{tabular}
		}
	\end{center}	\vspace{-10pt}
	\caption{Comparison of representative CPS anomaly detection approaches}
	\label{relatedwork::detectionmethods} \vspace{-10pt}
\end{table*}

\begin{itemize}[leftmargin=*]
	
\item \textit{Program behavior model.} Regarding the CPS anomaly detection based on program behavior models in the cyber domain, Yoon~\etal\cite{CoRR:YoonMCCS15} proposed a lightweight method for detecting anomalous executions using the distribution of system call frequencies. The frequencies are for individual system calls, \ie, 1-grams. The authors in~\cite{Abad:2013:CPSNA} proposed a hardware based approach for control-flow graph (CFG) validation in runtime embedded systems. McLaughlin~\etal \cite{McLaughlin:2014:NDSS} presented the Trusted Safety Verifier (TSV) to verify safety-critical code executed on programmable controllers, such as checking safety properties like range violations and interlocks of PLC programs.	
%Abera~\etal proposed the C-FLAT~\cite{Abera:CFLAT:2016}, a technique for remote attestation of the execution path of an application running on an embedded device.
C-FLAT~\cite{Abera:CFLAT:2016} instruments target control programs to achieve the remote attestation of execution paths of monitored programs, and the validity of control flow paths is based on static analysis.
Given an aggregated authenticator (\ie, fingerprint) of the program's control flow computed by the prover, the verifier is able to trace the exact execution path and thus can determine whether application's control 
flow has been compromised. C-FLAT~\cite{Abera:CFLAT:2016} is the most related work to our approach. Both C-FLAT and \ProjectName{} target at designing a general approach for detecting anomalous executions of embedded systems software. However, C-FLAT is insufficient to detect data-oriented attacks due to the lack of runtime execution context checking. 
It can only partially detect control intensity attacks with the assumption of knowing legal measurements of the target program. However, if the legal measurement covers a large range of sensor values, attacks can easily evade its detection because it does not check runtime consistency between program behavior and physical context. 

%C-FLAT~\cite{Abera:CFLAT:2016} is the most related work to our approach. Both C-FLAT and \EFSA{} target at designing a general approach rather than for a specific CPS application.
	
\item \textit{Traffic-based model.} Control systems exhibit relatively simpler network dynamics compared with traditional IT systems, \eg, fixed network topology, regular communication patterns, and a limited number of communication protocols. As a result, implementing network-based anomaly detection systems would be easier than traditional mechanisms. Feng~\etal\cite{Feng:2017:DSN} presented an anomaly detection method for ICS by taking advantage of the predictable and regular nature of communication patterns that exist between field devices in ICS networks. In the training phase, a base-line signature database for general packages is constructed. In the monitoring phase, the authors utilize Long Short-Term Memory (LSTM) network based softmax classifier to predict the most likely package signatures that are likely to occur given previously seen package traffic. The anomaly detector captures traffic anomalies if a package's signature is not within the predicted top $k$ most probable signatures according to the LSTM-based model. 
	
\item \textit{Timing-based model.} Several studies utilized timing information as a side channel to detect malicious intrusions. The rationale is that execution timing information is considered an important constraint for real-time CPS applications, and mimicking timing is  more difficult than mimicking the execution sequence. To this end, Zimmer~\etal\cite{Zimmer:2010:TID} used the worst-case execution time (WCET) obtained through static analysis to detect code injection attacks in CPS. Such timing-based detection technique is realized by instrumenting checkpoints within real-time applications. Sibin~\etal\cite{Mohan:2013:SSS} focused on detecting intrusions in real-time control systems. Yoon~\etal\cite{Yoon:2013:RTAS} presented SecureCore, a multicore architecture using the timing distribution property of each code block to detect malicious activities in real-time embedded system. Lu~\etal\cite{Lu:2015:ACF} investigated how to reduce timing checkpoints without sacrificing detection accuracy in embedded systems.

\item \textcolor{clmark}{\textit{Redundant execution analysis.} Munir~\etal\cite{Munir:TDSC:2018} presented an integrated approach for the
design of secure and dependable automotive CPS by conducting the steer-by-wire (SBW) case study. To provide fault tolerance (FT) to SBW applications, they proposed the FT-RMT (redundant multi-threading) scheme, which executes
safety-critical computations on redundant threads and
detects an error if observing a
mismatch between the two threads' output. However, introducing redundant controllers into CPS incurs high cost such as increased code size and reduced performance. In addition, under our threat model, attackers may be able to compromise redundant controllers to evade detection.}
	
\item \textit{Range based model.} Enforcing data ranges is the simplest method to detect CPS anomalies in the physical domain. As long as sensor readings are outside a pre-specified normal range, the anomaly detector raises an alarm. Hadziosmanovic~\etal \cite{Hadziosmanovic:2014} presented a non-obtrusive security monitoring system by deriving models for PLC variables from network packets as the basis for assessing CPS behaviors. For constant and attribute series, the proposed detection approach raises an alert if a value reaches outside of the enumeration set. However, range-based detection suffers from a low detection rate because it neglects the program's execution context, \eg, if the legal measurement covers a large range of sensor values, attacks can easily evade its detection. 

\item \textcolor{clmark}{\textit{Sensor and process noise fingerprinting.} 
Ahmed~\etal\cite{Ahmed:2018:NAD} proposed the NoisePrint, a CPS attack detection method based on process noise patterns (\eg, fluid sloshing in a storage tank) of the system. 
The intuition behind NoisePrint is that, sensor and process noise variations exhibit unique patterns among different processes. 
Therefore, it is hard for attackers to reproduce or control these noise variations, making them an ideal side information for CPS anomaly detection.
In the training phase, a combined
fingerprint dictionary for sensor and process noise is built under regular operations. Under sensor spoofing attack, noise
pattern deviates from the fingerprinted pattern is considered an anomaly. However, this approach requires an accurate noise measurement, which is subject to the ambient background noise in a process plant.}
	
\item \textit{Physical laws.} The idea of using physical models to define normal operations for anomaly detection is that, system states must follow immutable laws of physics.	Wang~\etal \cite{Wang:2014:ESORICS} derived a graph model to defeat false data injection attacks in SCADA system. It captures internal relations among system variables and physical states. Cho~\etal \cite{Cho:2015:CAC} presented a brake anomaly detection system, which compares the brake data with the norm model to detect any vehicle misbehavior (\eg, due to software bugs or hardware glitches) in the Brake-by-Wire system. Other examples include utilizing fluid dynamics and electromagnetics as the basic laws to create prediction models for water system~\cite{Hadziosmanovic:2014} and power grid~\cite{Liu:2009:FDI}, respectively. Based on the prediction models and predefined threat constraints, these methods check whether sensor readings are consistent with the expected behaviors of a control system. Cardenas~\etal \cite{Cardenas:2011:AAP} proposed a physical model based detection method by monitoring the physical system under control, and the sensor and actuator values. The authors also proposed automatic response mechanisms by estimating the system states. Urbina~\etal \cite{Urbina:2016:LIS} discussed the limitations of existing physics-based attack detection approaches, \ie, they cannot limit the impact of  stealthy attacks. The authors proposed a metric to measure the impact of stealthy attacks and to study the effectiveness of physics-based  detection.
	
\item \textit{Control policies.} Physical model can also be specified by control policies. The main purpose of the policies is to improve the survivability of control systems, \ie, without losing critical functions under attacks. For example, McLaughlin~\etal \cite{McLaughlin:2013:CSP} introduced a policy enforcement for governing the usage of CPS devices, which checks whether the policy allows an operation depending on the state of the plant around the time the operation was issued. The policies specify what behaviors should be allowed to ensure the safety of physical machinery and assets. 
	%We need to design new attack-resilient algorithms and architectures. 
	%The idea is to model how the output sequence (\ie, actuator values) of the physical system should react to the control input sequence (\ie, sensor values). 
	
\item \textit{Cyber-physical model.} Such a model captures the cyber-physical context dependency of control programs. Our proposed \EFSA{} characterizes control-program behaviors with respect to events, and enforces the runtime consistency among control decisions, values of data variables in control programs, and the physical environments. Thus, it is able to detect inconsistencies between the physical context and program execution. 

	%\index{event}
\end{itemize}

%here a summary 

As shown in Table~\ref{relatedwork::detectionmethods}, cyber models and physical models have different security guarantees. The former targets at detecting CPS control program anomalies in the cyber domain. While the latter mainly focuses on detecting false data injection attacks in the physical domain~\cite{Liu:2009:FDI}. 
The cyber-physical interaction (\ie, interactions between cyber components and physical components) in CPS makes it challenging to predict runtime program behaviors through static analysis of the program code or model training. Existing cyber models~\cite{Abera:CFLAT:2016, CoRR:YoonMCCS15} are effective against control-oriented attacks, however, insufficient to detect data-oriented attacks. An effective CPS program anomaly detection needs to reason about program behaviors with respect to cyber-physical interactions, \eg, the decision of opening a valve has to be made based on the current water level of the tank.
ContexIoT~\cite{jia2017contexiot} provides context identification for sensitive actions in the permission granting process of IoT applications on Android platforms. Though both ContextIoT and \EFSA{} consider execution contextual integrity, ContextIoT does not support the detection of data-oriented attacks.

Distinctive from existing works in this area, our \emph{Orpheus} focuses on utilizing the event-driven feature in control-program anomaly detection and our program behavior model combines both the cyber and physical aspects. Consequently, physics-based models, which can be inherently integrated into our approach to enhance security and efficiency, do not compete but rather complement our scheme. Stuxnet attack~\cite{stuxnet:2013} manipulated the nuclear centrifuge's rotor speed, and fooled the system operator by replaying the recorded normal data stream during the attack~\cite{Physics:NDSS:2017}.
Since \EFSA{}'s detection is independent on the history data, it makes Stuxnet-like attacks detectable in \EFSA{} by detecting runtime inconsistencies between the physical context (runtime rotor speed) and the control program's behavior.
In addition, attackers may exploit hardware vulnerabilities~\cite{vanderVeen:2016:DDR} to manipulate data in memory so as to launch attacks on control branch or control intensity. \EFSA{} is also able to detect anomalies caused by such hardware attacks.

\section{Conclusion}\label{sec:Conclusion}
In this work, we presented \emph{Orpheus}, a new security mechanism for CPS control programs in defending against data-oriented attacks, by enforcing cyber-physical execution semantics. As an \FSA{}-based instantiation of \emph{Orpheus}, we proposed the program behavior model \EFSA{}, which advances the state-of-the-art program behavior modelling. 
To the best of our knowledge, this is the first program behavior model that integrates both cyber and physical properties to defend against data-oriented attacks. We implemented a proof-of-concept prototype to demonstrate the feasibility of our approach. Real-world case studies demonstrated \EFSA{}'s efficacy against different data-oriented attacks. As for our future work, we plan to integrate physics-based models into our approach, design robust event verification mechanisms, and extend the \emph{Orpheus} design paradigm to support actuation integrity for fine-grained anomaly detection at the instruction level without the need of tracing facilities. 
We also plan to investigate the scalability of our approach on program size and complexity. 

%[us rephrase] evertheless, our proposed approach is generic and it can be applied on different CPSs with various real-time constraints. In the future, we intend to conduct further experiments on power grid and robotics sys- tems.

%to check the  Open-SWaT and Open-SecUTS  code???

% use section* for acknowledgment
\ifCLASSOPTIONcompsoc
  % The Computer Society usually uses the plural form
  \section*{Acknowledgments}
\else
  % regular IEEE prefers the singular form
  \section*{Acknowledgment}
\fi

This work has been supported by the Office of Naval Research under Grant ONR-N00014-17-1-2498, National Science Foundation under Grant OAC-1541105, and Security and Software Engineering Research Center (S2ERC), a NSF sponsored multi-university Industry/University Cooperative Research Center (I/UCRC).

% Can use something like this to put references on a page
% by themselves when using endfloat and the captionsoff option.
\ifCLASSOPTIONcaptionsoff
  \newpage
\fi

% trigger a \newpage just before the given reference
% number - used to balance the columns on the last page
% adjust value as needed - may need to be readjusted if
% the document is modified later
%\IEEEtriggeratref{8}
% The "triggered" command can be changed if desired:
%\IEEEtriggercmd{\enlargethispage{-5in}}

% references section

% can use a bibliography generated by BibTeX as a .bbl file
% BibTeX documentation can be easily obtained at:
% http://mirror.ctan.org/biblio/bibtex/contrib/doc/
% The IEEEtran BibTeX style support page is at:
% http://www.michaelshell.org/tex/ieeetran/bibtex/
%\bibliographystyle{IEEEtran}
% argument is your BibTeX string definitions and bibliography database(s)
%\bibliography{IEEEabrv,../bib/paper}
%
% <OR> manually copy in the resultant .bbl file
% set second argument of \begin to the number of references
% (used to reserve space for the reference number labels box)
\bibliographystyle{IEEEtran}
\bibliography{ref} 

% biography section
% 
% If you have an EPS/PDF photo (graphicx package needed) extra braces are
% needed around the contents of the optional argument to biography to prevent
% the LaTeX parser from getting confused when it sees the complicated
% \includegraphics command within an optional argument. (You could create
% your own custom macro containing the \includegraphics command to make things
% simpler here.)
%\begin{IEEEbiography}[{\includegraphics[width=1in,height=1.25in,clip,keepaspectratio]{mshell}}]{Michael Shell}
% or if you just want to reserve a space for a photo:

%\begin{IEEEbiography}{Michael Shell} clip keepaspectratio
	\begin{IEEEbiographynophoto}
%\begin{IEEEbiography}[{\includegraphics[width=1.09in,height=3cm,]{longcheng.pdf}}]
{Long Cheng} is currently an assistant professor of computer science at Clemson University, USA. He received his second PhD in Computer Science from Virginia Tech USA in 2018, and the first PhD from Beijing University of Posts and Telecommunications China in 2012. He worked as a Research Scientist at the Institute for Infocomm Research (I$^2$R), Singapore from 2014 to 2015, and a Research Fellow at Singapore University of Technology and Design from 2012 to 2014. His research interests include system and network security, cyber forensics, cyber-physical systems (CPS) security, Internet of Things, mobile computing, and wireless networks. He received the Best Paper Award from IEEE Wireless Communications and Networking Conference (WCNC) in 2013, the Erasmus Mundus Scholar Award in 2014, and the Pratt Fellowship at Virginia Tech in 2017.

\end{IEEEbiographynophoto}
%\end{IEEEbiography}
%\balance

%\begin{IEEEbiography}[{\includegraphics[width=1.0in,height=3cm,]{ke.eps}}]
	\begin{IEEEbiographynophoto}
	{Ke Tian} is currently a data scientist at Microsoft. His experience is on leveraging Machine Learning and Anomaly Detection to address cyber security challenges. He received his PhD degree from the Department of Computer Science at Virginia Tech in 2018. He received his bachelor degree majoring information security from University of Science and Technology of China in 2013. His research interests are in cybersecurity, mobile security and machine learning.   
%\end{IEEEbiography}
\end{IEEEbiographynophoto}

% if you will not have a photo at all:
%\begin{IEEEbiography}[{\includegraphics[width=0.98in,height=3cm,]{dan.eps}}]
\begin{IEEEbiographynophoto}
	{Danfeng (Daphne) Yao} is an associate professor of computer science at Virginia Tech. In the past decade, she has been working on designing and developing data-driven anomaly detection techniques for securing networked systems against stealthy exploits and attacks. Her expertise also includes mobile security. Dr. Yao received her Ph.D. in Computer Science from Brown University. Dr. Yao is an Elizabeth and James E. Turner Jr. '56 Faculty Fellow and L-3 Faculty Fellow. She received the NSF CAREER Award in 2010 for her work on human-behavior driven malware detection, and the ARO Young Investigator Award for her semantic reasoning for mission-oriented security work in 2014. She has several Best Paper Awards (e.g., ICNP '12, CollaborateCom '09, and ICICS '06) and Best Poster Awards (e.g., ACM CODASPY '15). She was given the Award for Technological Innovation from Brown University in 2006. She held multiple U.S. patents for her anomaly detection technologies. Dr. Yao is an associate editor of IEEE Transactions on Dependable and Secure Computing (TDSC). She serves as PC members in numerous computer security conferences, including ACM CCS and IEEE S\&P. She has over 85 peer-reviewed publications in major security and privacy conferences and journals.
	\end{IEEEbiographynophoto}
%\end{IEEEbiography}

% insert where needed to balance the two columns on the last page with
% biographies
%\newpage

%\begin{IEEEbiography}[{\includegraphics[width=1.0in,height=3cm,]{luisha.eps}}]
\begin{IEEEbiographynophoto}
	{Lui Sha} received the Ph.D. degree in computer science from Carnegie Mellon University, Pittsburgh, PA, USA, in 1985. He is currently a Donald B. Gillies Chair Professor of Computer Science with the University of Illinois at Urbana Champaign, Champaign, IL, USA. His work on real-time computing is supported by most of the open standards in real-time computing and has been cited as a key element in the success of many national high-technology projects including GPS upgrade, the Mars Pathfinder, and the International Space Station.
\end{IEEEbiographynophoto}
%\end{IEEEbiography}

%\begin{IEEEbiography}[{\includegraphics[width=1.1in,height=3cm,]{Raheem.eps}}]
\begin{IEEEbiographynophoto}
	{Raheem Beyah} is the Motorola Foundation Professor and Associate Chair for Strategic Initiatives and Innovation in the School of Electrical and Computer Engineering at Georgia Tech, where he leads the Communications Assurance and Performance Group (CAP). His research interests include network security, wireless networks, network traffic characterization and performance, and security visualization. He received the National Science Foundation CAREER award in 2009 and was selected for DARPA's Computer Science Study Panel in 2010. He is a member of AAAS, ASEE, a lifetime member of NSBE, a senior member of IEEE, and an ACM Distinguished Scientist.
\end{IEEEbiographynophoto}
%\end{IEEEbiography}
% You can push biographies down or up by placing
% a \vfill before or after them. The appropriate
% use of \vfill depends on what kind of text is
% on the last page and whether or not the columns
% are being equalized.

%\vfill

% Can be used to pull up biographies so that the bottom of the last one
% is flush with the other column.
%\enlargethispage{-5in}

% that's all folks
\end{document}